\documentclass[aps,prb,twocolumn, amsmath, amssymb,showpacs,superscriptaddress]{revtex4-1}

\usepackage{graphicx}
\usepackage{amsmath,amssymb,amsfonts}
\usepackage{textcomp}
\usepackage{hyperref}
\usepackage{gensymb}
\usepackage{verbatim}
\usepackage{amsmath}
\hypersetup{breaklinks=true,colorlinks=true,urlcolor=black}
\usepackage{color}
\usepackage{soul}
\DeclareGraphicsExtensions{.jpg,.pdf,.png,.eps}

\begin{document}

\title{Magnetism and $T-x$ phase diagrams of Na and Ag substituted EuCd$_2$As$_2$}
\author{Brinda Kuthanazhi}
\affiliation{Ames National Laboratory, Iowa State University, Ames, Iowa-50011, USA}
\affiliation{Department of Physics and Astronomy, Iowa State University, Ames, Iowa-50011, USA}
\author{Kamal R. Joshi}
\affiliation{Ames National Laboratory, Iowa State University, Ames, Iowa-50011, USA}
\affiliation{Department of Physics and Astronomy, Iowa State University, Ames, Iowa-50011, USA}
\author{Sunil Ghimire}
\affiliation{Ames National Laboratory, Iowa State University, Ames, Iowa-50011, USA}
\affiliation{Department of Physics and Astronomy, Iowa State University, Ames, Iowa-50011, USA}
\author{Erik Timmons}
\affiliation{Ames National Laboratory, Iowa State University, Ames, Iowa-50011, USA}
\affiliation{Department of Physics and Astronomy, Iowa State University, Ames, Iowa-50011, USA}
\author{Lin-Lin Wang}
\affiliation{Ames National Laboratory, Iowa State University, Ames, Iowa-50011, USA}
\author{Elena Gati}
\altaffiliation[Current address: ]{Max Planck Institute for Chemical Physics of Solids, 01187 Dresden, Germany}
\affiliation{Ames National Laboratory, Iowa State University, Ames, Iowa-50011, USA}
\affiliation{Department of Physics and Astronomy, Iowa State University, Ames, Iowa-50011, USA}
\author{Li Xiang}
\altaffiliation[Current address: ]{National High Magnetic Field Laboratory, Florida State University, Tallahassee, Florida 32310, USA}
\affiliation{Ames National Laboratory, Iowa State University, Ames, Iowa-50011, USA}
\affiliation{Department of Physics and Astronomy, Iowa State University, Ames, Iowa-50011, USA}
\author{Ruslan Prozorov}
\affiliation{Ames National Laboratory, Iowa State University, Ames, Iowa-50011, USA}
\affiliation{Department of Physics and Astronomy, Iowa State University, Ames, Iowa-50011, USA}
\author{Sergey L. Bud'ko}
\affiliation{Ames National Laboratory, Iowa State University, Ames, Iowa-50011, USA}
\affiliation{Department of Physics and Astronomy, Iowa State University, Ames, Iowa-50011, USA}
\author{Paul C. Canfield}
\email{canfield@ameslab.gov}
\affiliation{Ames National Laboratory, Iowa State University, Ames, Iowa-50011, USA}
\affiliation{Department of Physics and Astronomy, Iowa State University, Ames, Iowa-50011, USA}

\date{\today}

\begin{abstract}
EuCd$_2$As$_2$ is an antiferromagnetic semimetal, that can host non-trivial topological properties, depending upon its magnetic state and excitations. Here, we report the synthesis and characterization of Eu(Cd$_{1-x}$Ag$_x$)$_2$As$_2$ and Eu$_{1-y}$Na$_y$Cd$_2$As$_2$, and study the evolution and nature of magnetic order with doping. Temperature-substitution phase diagrams are constructed from the electrical resistance and magnetic susceptibility data. We observe a splitting of the magnetic transition into two different transitions, and the gradual increase in one of the transition temperatures with Ag- and Na-substitution. The other transition remains more or less independent of doping. We further show that a magnetic state with a net ferromagnetic moment is stabilized by both Ag and Na doping and this can be explained by considering the changes in band filling due to substitution as suggested by density functional theory (DFT) calculations. We thus show that chemical substitution and the subsequent changes in band filling could be a pathway to tune the magnetic ground state  and to stabilize a ferromagnetic phase in EuCd$_2$As$_2$.
\end{abstract}

\maketitle

\section{Introduction}
The search for new topological materials and the realization of novel surface states and magneto-transport phenomena has driven condensed matter research in the recent years. \cite{Hsieh2008, Hasan2010, Burkov2011, Burkov2016, Chiu2016, Felser2017, Armitage2018, Tokura2019} One such material class is Weyl semimetals, which are expected to show a chiral anomaly, large anomalous Hall effect, Fermi arc surface states, unconventional optical response etc. \cite{Wan2011, Parameswaran2014, Chan2016, Jia2016, Chan2017, Shekhar2018} Weyl semimetals can be realized from a Dirac semi-metal, which hosts linear band dispersion at certain symmetry protected Dirac points. Breaking either time reversal or inversion symmetry lifts the degeneracy of Dirac points, giving rise to a Weyl semi-metal. A minimum number of Weyl points can be attained by breaking the time reversal symmetry, \cite{Armitage2018} which can be achieved through magnetic ordering. Despite a large number of materials being identified as magnetic Weyl semimetal candidates, unambiguous experimental signatures, especially in magneto-transport, have remained difficult to attain and challenging to untangle from other contributions. One among the major road blocks for this is the presence of other topologically trivial bands close to the Fermi level, \cite{Ramshaw2018} combined with a large number of Weyl points usually observed. \cite{Weng2015, Huang2015, Arnold2016, Chang2018} It is in this context that Weyl semimetal candidate EuCd$_2$As$_2$ garnered a lot of attention, as DFT calculations predicted it to host a single pair of Weyl points, when stabilized in a ferromagnetic state with moments along the crystallographic $c-$direction. \cite{Hua2018, Wang2019} 

Subsequent experimental work showed direct and indirect signatures of Weyl physics, as well as the possibility of this system being tunable between various magnetic ground states. \cite{Rahn2018, Ma2019, Niu2019, Jo2020, Ma2020, Taddei2022, Sanjeewa2020, Gati2021, Xu2021} Early experiments on EuCd$_2$As$_2$ reported it to crystallize in the hexagonal space group $P\bar{3}m1$ with an antiferromagnetic (AFM) transition around $10~$K. \cite{Artmann1996, Schellenberg2010} Later, the magnetic ground state was reported to be AFM-A, where moments are aligned ferromagnetically (FM) in-plane and antiferromagnetically inter-plane. \cite{Rahn2018} Whereas a c-polarized ferromagnetic state was predicted to be favorable for a single pair of Weyl points, other magnetic ground states were still expected to show topologically non-trivial states. \cite{Wang2019, Jo2020} Observation of anomalous Hall effect in transport measurements, \cite{Soh2019} soon followed by the direct observation of Weyl points by angle resolved photo emission spectroscopy (ARPES) in the paramagnetic state have confirmed this prediction. \cite{Ma2019} This was followed by reports of a modified synthesis procedure stabilizing a ferromagnetic ground state, \cite{Jo2020} which was later reproduced and the FM ground state confirmed by neutron diffraction. \cite{Taddei2022} Furthermore, Ba-substitution in EuCd$_2$As$_2$, has also shown an FM state with spin canting. \cite{Sanjeewa2020} Hydrostatic pressure has also been used to tune the magnetic ground state revealing that pressures as low as $2~$GPa can change the ambient-pressure, AFM ordering of pure EuCd$_2$As$_2$ into an ordered state with a net ferromagnetic component with moments in-plane. \cite{Gati2021, Yu2022, Du2022} It is also known to go through a metallic - insulating - metallic transition below the magnetic transition with applied pressures. \cite{Du2022} A recent study suggests that higher pressures ($\sim19~$GPa) turn the magnetic ordering in to ferromagnetic with moments out of plane. \cite{Yu2022}  

In this paper, we further probe the effect of chemical tunability on magnetism, by careful chemical substitution of Ag and Na on Cd and Eu sites respectively of EuCd$_2$As$_2$. Electrical resistivity and magnetic susceptibility measurement results are presented, and the temperature-substitution phase diagrams are constructed from them. Furthermore, we use magneto-optical Kerr effect measurements to further support the conclusion that whereas pure EuCd$_2$As$_2$ is in an AFM state, Na and Ag substituted EuCd$_2$As$_2$ has a measurable FM component, even for the smallest doping we have. Finally, we also discuss a possible mechanism that is leading to stabilization of different magnetic states, supported by DFT calculations.

\begin{table}
	\begin{tabular}{c|c|c}
		\hline\hline
		Sample & Nominal $x$, $y$ & Measured $x$, $y$\\
		\hline
		Eu(Cd$_{1-x}$Ag$_x$)$_2$As$_2$& $x= 0$ & $x = 0$ \\
		&$x = 0.05$ &$x = 0.0008(1)$\\
		&$x = 0.10$ &$x = 0.0019(3)$\\
		&$x = 0.20$ &$x = 0.0040(3)$\\
		&$x = 0.30$ &$x = 0.0055(3)$\\
		\hline 
		Eu$_{1-y}$Na$_y$Cd$_2$As$_2$&$y=0$&$y=0$\\
		&$y = 0.10$&$y = 0.005(1)$\\
		&$y = 0.20$&$y = 0.008(1)$\\
		\hline\hline
	\end{tabular}
	\caption{The nominal and measured concentrations of Ag and Na in various substituted EuCd$_2$As$_2$ crystals.}
	\label{xwds}
\end{table}

\begin{figure*}
	\includegraphics[scale=1]{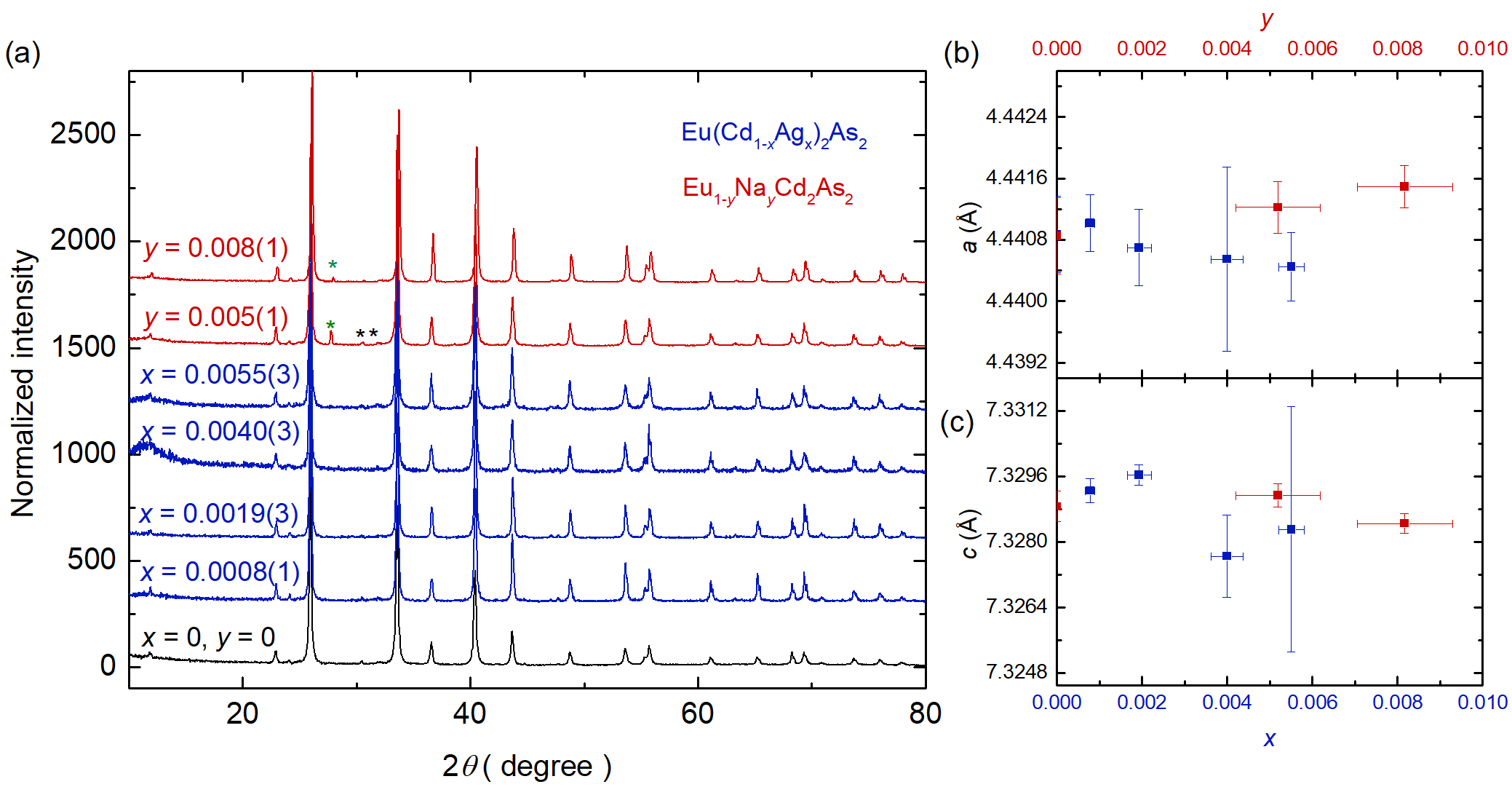}
	\caption{(a) Powder x-ray diffraction patterns of various Ag and Na substituted EuCd$_2$As$_2$ samples. The intensities have been normalized such that the strongest peak (the (101) peak located at $2\theta \sim$ $26~$degrees) has 1000 counts. The impurity peaks marked by black and green stars are from Sn and NaSn$_2$As$_2$ respectively. Each data set is offset by a constant to show the diffraction patterns clearly. (b) Lattice parameter $a$ and (c) $c$ as a function of Ag-substitution $x$ and Na-substitution $y$.}
	\label{pxrd}
\end{figure*}

\section{Methods}

The single crystals of pure and Na- and Ag-substituted EuCd$_2$As$_2$ were grown out of Sn flux with following procedures. For pure EuCd$_2$As$_2$ an initial stoichiometry of Eu:Cd:As:Sn = 1:2:2:10 (Eu - Ames Laboratory, 99.99+ \%, Cd - Alfa Aesar, 99.9997 \%, As - Alfa Aesar, 99.9999 \%, and Sn - Alfa Aesar, 99.99\% ) was put into a fritted alumina crucible (CCS) \cite{Canfield2016} and sealed in fused silica tube under a partial pressure of argon. The thus prepared ampoule was heated up to $900\degree$~C over 24 hours, and held there for 20 hours. This was followed by a slow cooling to $550\degree$~C over 200 hours, and decanting of the excess flux using a centrifuge. \cite{Canfield1992, Canfield2019} 
For the synthesis of Ag and Na substituted samples, the initial composition was modified as; Eu$_1$(Cd$_{1-x}$Ag$_x$)$_2$As$_2$Sn$_{10}$ for $x = 0.05, 0.10, 0.20, 0.30$ and Eu$_{1-y}$Na$_y$Cd$_2$As$_2$Sn$_{10}$ for $y=0.10, 0.20$ respectively. Further increasing the Ag and Na content in the initial melt resulted in secondary phase inclusions appearing in the crystal. 
\begin{figure*}
	\includegraphics[scale=1]{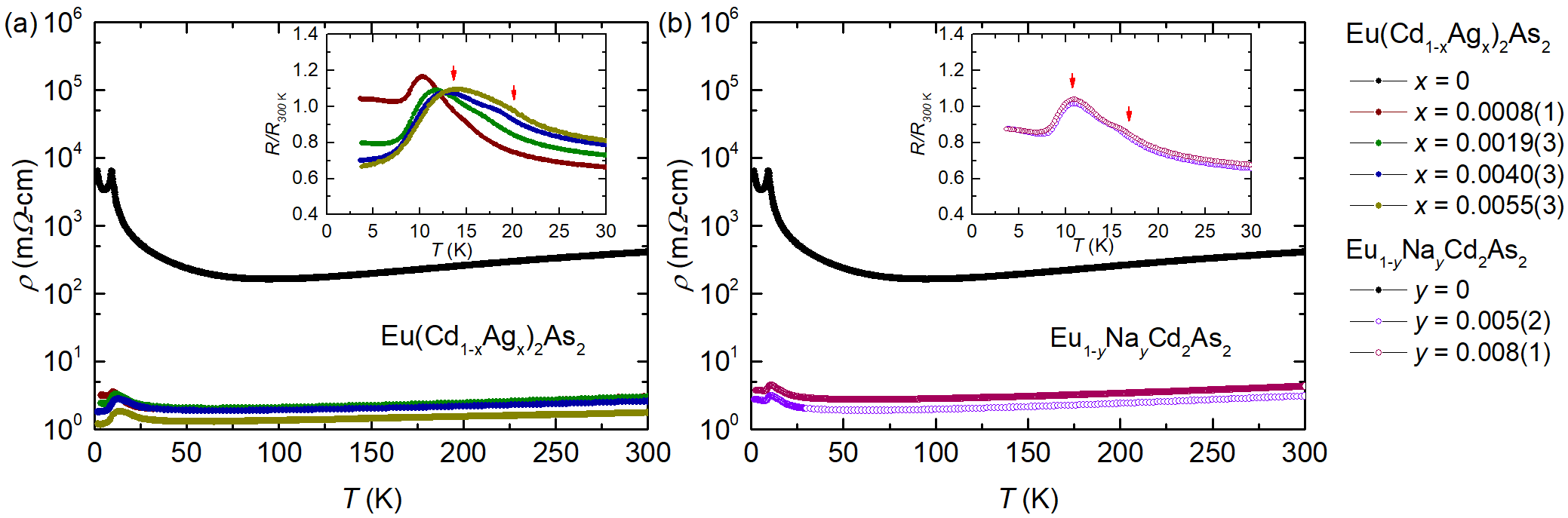}
	\caption{Resistivity as a function of temperature, $\rho(T)$, plotted on a semi-log scale for undoped EuCd$_2$As$_2$ and various (a) Ag doped samples and (b) Na doped samples. The undoped sample has two orders of magnitude higher resistivity than the doped samples. It also has a significant upturn above the magnetic transition, whereas this feature is less pronounced in all substituted samples. Inset shows the low temperature regime for the doped samples. Two features are observed for samples with higher doping, illustrated by two red arrows for $x=0.0055(3)$ and $y=0.008(1)$ samples.} 
	\label{rt}
\end{figure*}
\begin{figure*}
	\includegraphics[scale=1]{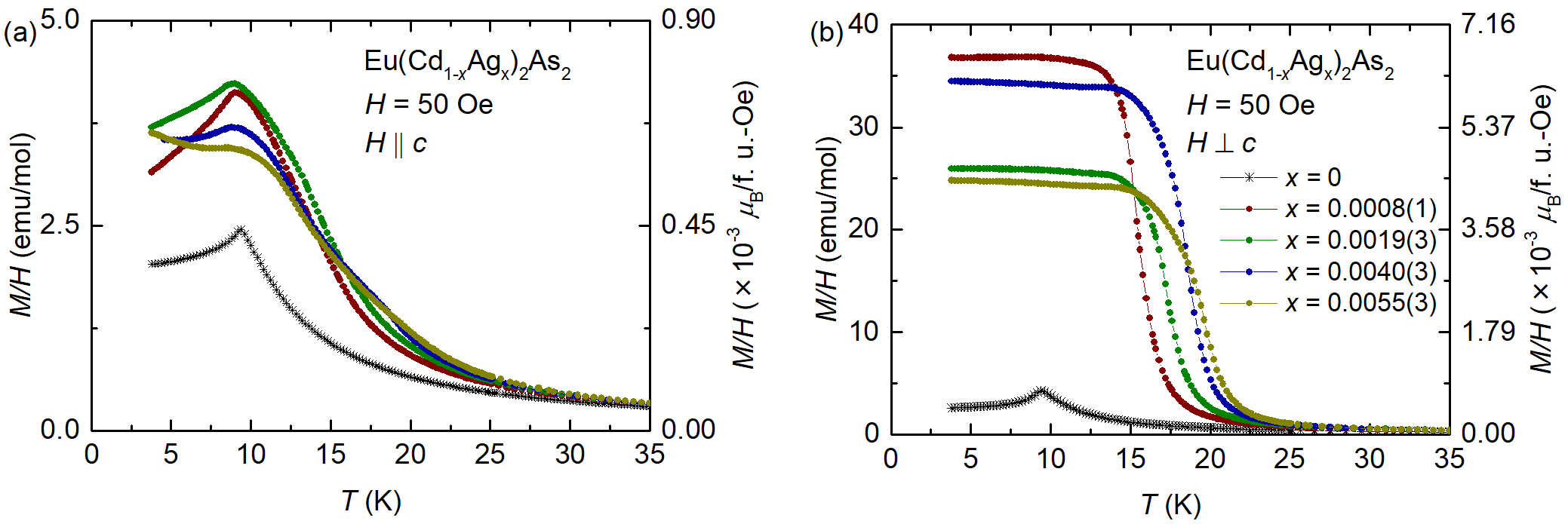}
	\caption{Temperature dependent magnetization divided by applied field, $M(T)/H$, as a function of temperature for various Ag substituted samples with an applied field of $H=50~$Oe (a) parallel to crystallographic $c$ direction and (b) in the $ab-$plane. The data shown is above $T=3.8~$K, as there were slight features below that, due to presence of Sn (flux) in the sample which goes superconducting below this temperature. Right $y-$ axis shows the scale in units of $\mu_B/$f. u.-Oe.}
	\label{mt_Ag}
\end{figure*}
\begin{figure*}
	\includegraphics[scale=1]{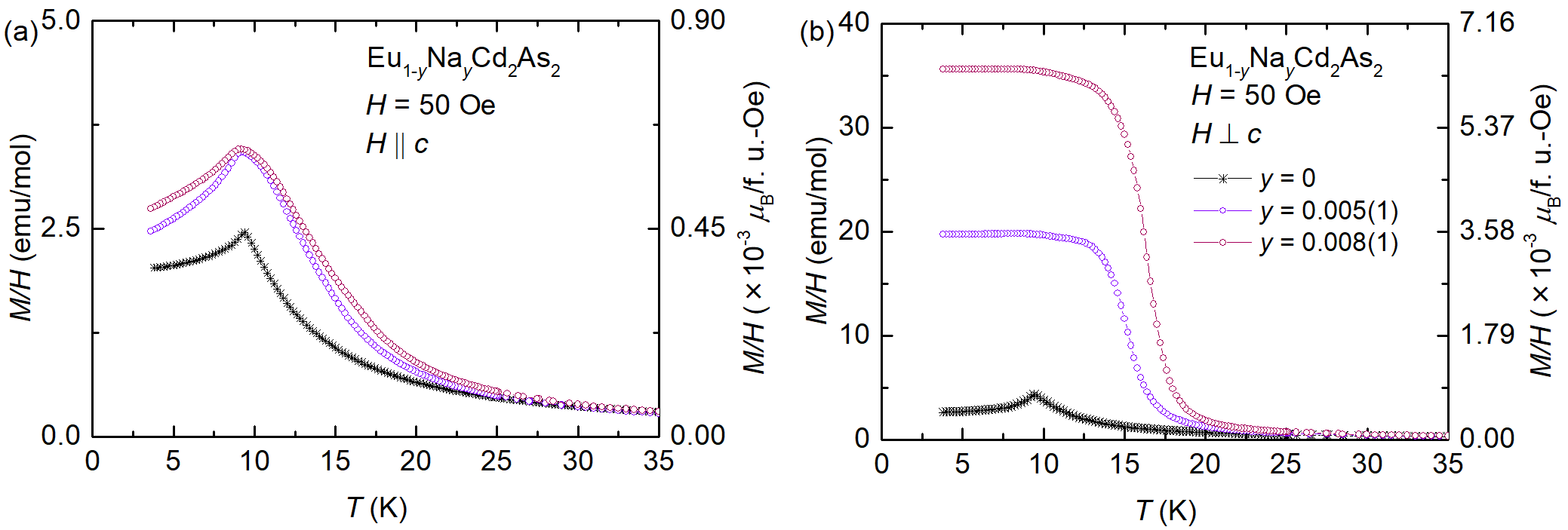}
	\caption{Temperature dependent magnetization divided by applied field, $M(T)/H$, as a function of temperature for various Na substituted samples with an applied field of $H=50~$Oe (a) parallel to crystallographic $c$ direction and (b) in the $ab-$plane. The data shown is above $T=3.8~$K, as there were slight features below that, due to presence of Sn (flux) in the sample which goes superconducting below this temperature. Right $y-$ axis shows the scale in units of $\mu_B/$f. u.-Oe.}
	\label{mt_Na}
\end{figure*}

Powder x-ray diffraction measurements were carried out to determine the phase purity and track the changes in lattice parameters. Each powder x-ray diffraction pattern was collected on ground crystals using a Rigaku Miniflex-II diffractometer, with Cu$K_{\alpha}$ radiation. Lattice parameters were obtained from a Rietveld refinement on it using GSAS-II software. \cite{Toby2013} 

The actual concentrations of Na and Ag in the grown crystals (as opposed to the compositions of the initial, high temperature melt), were determined using Energy dispersive X-ray Spectroscopy (EDS) and Wavelength dispersive X-ray Spectroscopy (WDS). The SEM was an FEI Quanta-FEG 250 outfitted with Oxford Instruments EDS and WDS. All was run under Aztec software, version 5.0. Initial experiments with EDS showed no discernible peaks for Ag or Na due to peak overlap with Cd and Eu, respectively. Hence, WDS analysis was used to quantify Na and Ag content, whereas other elements were determined with EDS. Na- and Ag-substituted EuCd$_2$As$_2$ samples were analyzed in the SEM at 20~kV, 8-16~nA, and 90-260 seconds of acquisition time. For Na-substituted samples, measurements were done on 1-4 different points each for two different crystals of each nominal substitution. The WDS analyses ran about 90 seconds each, to achieve 0.01 weight\% reproducibility. For crystals with various Ag substitutions, the samples were embedded and polished along the cross section for the best results. Samples were coated with 10 nm of carbon for conductivity. Wavelength scans over the Ag$-L_{\alpha}$ peak showed a small, but discernible peak. Samples were measured at a minimum of two points for each of the embedded crystals. The WDS analyses ran about 260 seconds each to attain 0.02\% reproducibility.  

The magnetic measurements were carried out in a Quantum Design, Magnetic Property Measurement System with applied fields up to 50 kOe. The samples were mounted on a Poly-Chloro-Tri-Fluoro-Ethylene disk, and the separately measured background of the disk was subtracted. The AC resistivity measurements were done in a Quantum Design, Physical Property Measurement System in the standard four point configuration. The current was passed in the $ab-$ plane with $I=3~$mA and $f=17~$Hz. Contacts were made using Epo-tek H20E silver epoxy. Pure EuCd$_2$As$_2$ crystals were very difficult to make low resistance contacts on, as the contact resistance remained several tens of Ohms despite best efforts, which makes the measurements challenging. As will be discussed below, pure EuCd$_2$As$_2$ has semiconducting- like resistivity values as well as low temperature behavior. Any finite substitution/doping level we add results in much lower contact resistances as well as lower resistivity values. A lot of the resistance data shown later is represented as normalized resistances for comparison between the samples and to get rid of the errors in geometrical factors. As soon as we substitute Ag or Na, the contact resistances values drops to 3-4 ohms. Along with this, the absolute value of measured resistance also drops by two orders of magnitude as we go from pure EuCd$_2$As$_2$ to any of the substituted sample.

Magneto optical Kerr effect measurements were carried out to confirm the existence of ferromagnetic domains. The off-diagonal components of anisotropic permittivity depend on magnetization vector amplitude and orientation. This leads to the anisotropy of the speed of light in different directions, causing linearly polarized light to rotate and become elliptically polarized. Depending on the angle of incidence and the direction of polarization, different components of the magnetization vector are engaged. We used the magneto-optical polar Kerr effect for a real-space 2D imaging of the magnetization component perpendicular to the reflection surface. 

In the experiment, linearly polarized light is incident perpendicular to the sample face, thus, it is only sensitive to the perpendicular component of the surface magnetization vector. Unlike Kerr spectroscopies, we use a polarized-light optical microscope where the transmitted through an analyzer light intensity reveals a real-space 2D image of the Kerr rotation angle, proportional to local magnetization amplitude. Up and down directions of surface magnetization cause rotation clockwise and counter-clockwise. With a slightly misaligned analyzer (1-3 degrees away from 90 degrees) this results in the optical contrast where “up” domains are lighter and “down” domains are darker. The imaging was performed with Olympus polarized microscope mounted above a closed cycle $^4$He cryostat (Montana Instruments) with the ability to apply a magnetic field along and perpendicular to the sample. The sample was mounted on a gold-plated copper cold finger with a dash of Apiezon grease. 

Total energy and density of states with spin-orbit coupling (SOC) in density functional theory \cite{Kohn1964, Sham1965} (DFT) have been calculated with PBE \cite{Perdew1996} exchange-correlation functional, a plane-wave basis set and projected augmented wave method \cite{Blochl1994} as implemented in VASP. \cite{Kresse1996, GKresse1996} To account for the half-filled strongly localized Eu 4f orbitals, a Hubbard-like \cite{Sutton1998} U is used. For A-type anti-ferromagnetic (AFMA) and ferromagnetic (FM) EuCd$_2$As$_2$, the hexagonal unit cell is doubled along the c-axis and calculated on a $\Gamma-$centered Monkhorst-Pack \cite{Pack1976} (11$\times$ 11$\times$ 3) $k-$point mesh with a Gaussian smearing of 0.05 eV. A kinetic energy cutoff of 254 eV and experimental lattice parameters \cite{Schellenberg2011} have been used with atoms fixed in their bulk positions. To calculate the energy difference between AFMA and FM with doping, the total energy of a charged cell with a uniform neutralizing background \cite{Makov1995} has been used.

\section{Results and Discussion}

\begin{figure*}
	\includegraphics[scale=0.8]{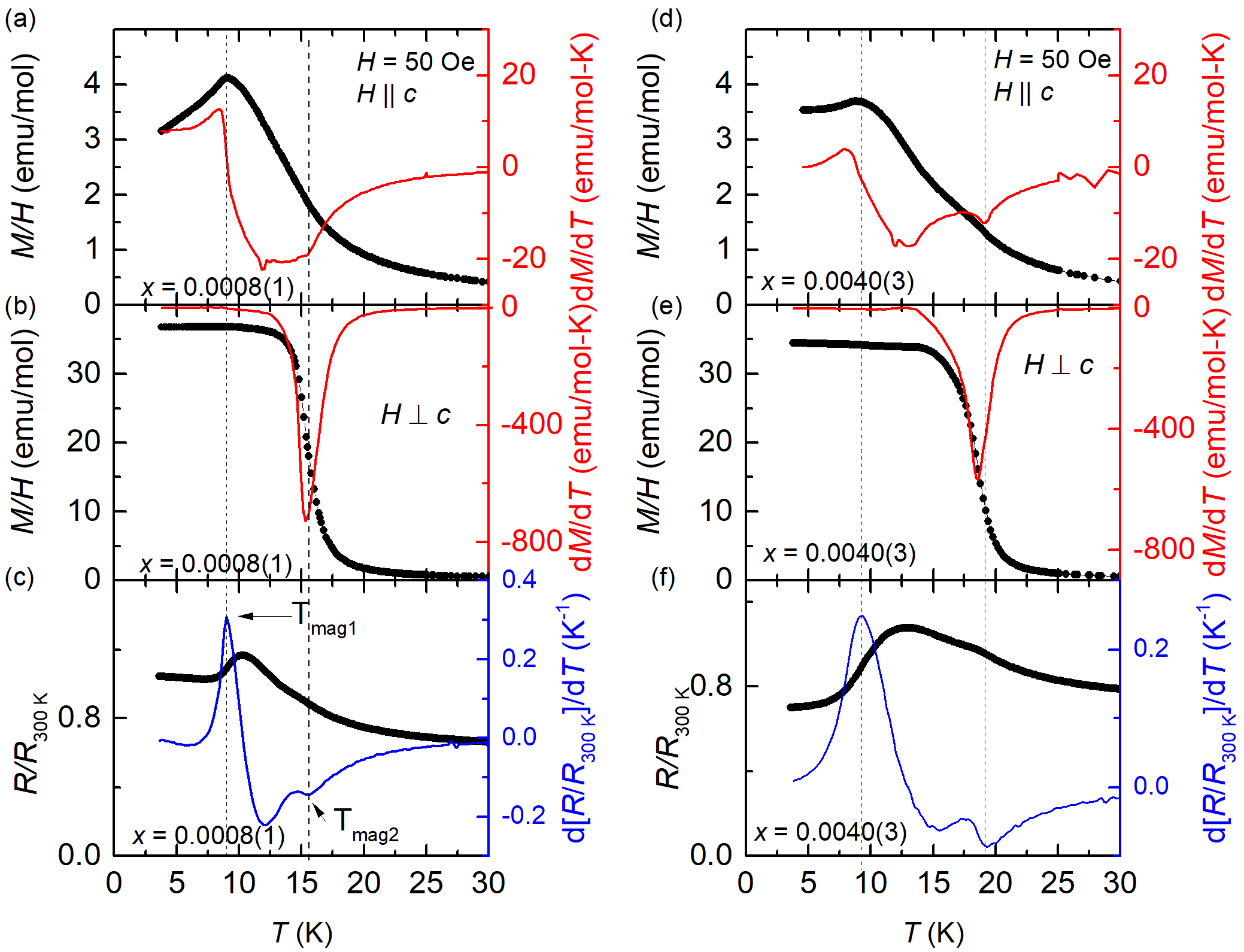}
	\caption{The low temperature dependences of normalized resistance $R/R_{300 K}$, the temperature derivative of the normalized resistance $d[R(T)/R_{300 K}]/dT$, temperature dependent magnetization divided by field $M(T)/H$, and the derivative $dM/dT$ plotted together for $T\leq30~$K for $x=0.0008(1)$ and $x=0.0040(3)$ Ag substituted samples. (a) $M(T)/H$ plotted along with the temperature derivative $dM/dT$ for $H = 50~$Oe field cooled in $H \parallel c$ direction and (b) $H \perp c$ direction for $x=0.0008(1)$ substitution. (c) $R/R_{300 K}$ and $d[R(T)/R_{300 K}]/dT$ for $x=0.0008(1)$ substitution. The peak positions in $d[R(T)/R_{300 K}]/dT$, more or less coincides with the feature in $M(T)/H$. Panels (d), (e), and (f) show similar data for $x=0.0040(3)$ sample. The low field $M(T)$ and zero field $R(T)$ have small Sn features, therefore we truncate data below $T_c$ of Sn. Dotted lines show the transition temperatures determined from $d[R(T)/R_{300 K}]/dT$.}
	\label{mtrt1}
\end{figure*}

\subsection{Structure and substitution}

The concentration of Ag and Na determined from WDS for various initial concentrations used for synthesis are given in Table \ref{xwds}. It is worth highlighting the extremely small amounts of Ag and Na that can be incorporated in to the EuCd$_2$As$_2$ system. Although we started the synthesis with nominal concentration of $10\%$ - $20\%$ for Na and $5-30\%$ for Ag, the actual concentration of dopants in the crystals, as determined from WDS are orders of magnitude smaller. With very small concentrations of Ag and Na going in, and the changes in carrier density (indicated by resistivity behavior as will be discussed later), we are using the words substitution and doping interchangeably from here on. For Ag substituted samples, there is a very small but discernible decrease in the amount of Cd obtained in EDS, which suggests that Ag is going into the Cd sites. No such trends were seen in Na doped samples, we are assuming it is going into the Eu sites. These data indicate that there is a rather narrow width of formation for Na and Ag to go into the EuCd$_2$As$_2$ structure. For many compounds such small doping levels would be irrelevant, but given EuCd$_2$As$_2$'s extreme sensitivity to small deviations from stoichiometry \cite{Jo2020, Sanjeewa2020, Taddei2022} probing even this narrow width of formation is necessary and revealing. 

Figure \ref{pxrd} (a) shows the powder x-ray diffraction data for various samples. This confirms the $P\bar{3}m1$ structure for all the substituted crystals with minimal change in peak position. The small impurity peaks, denoted by black and red stars in Fig. \ref{pxrd}, identified in some of the patterns are from Sn and NaSn$_2$As$_2$ impurity phases respectively. The lattice parameters $a$ and $c$, as determined by fitting the power x-ray diffraction data, are shown in Fig. \ref{pxrd} (b) and (c) respectively. This shows the changes in lattice parameters with Ag- and Na-substitution are minimal and within error bars, which would be consistent with very small amounts of Ag and Na content in the crystals.

\subsection{Temperature dependent magnetization and resistivity data and constructing the phase diagrams}

\begin{figure*}
	\includegraphics[scale=0.8]{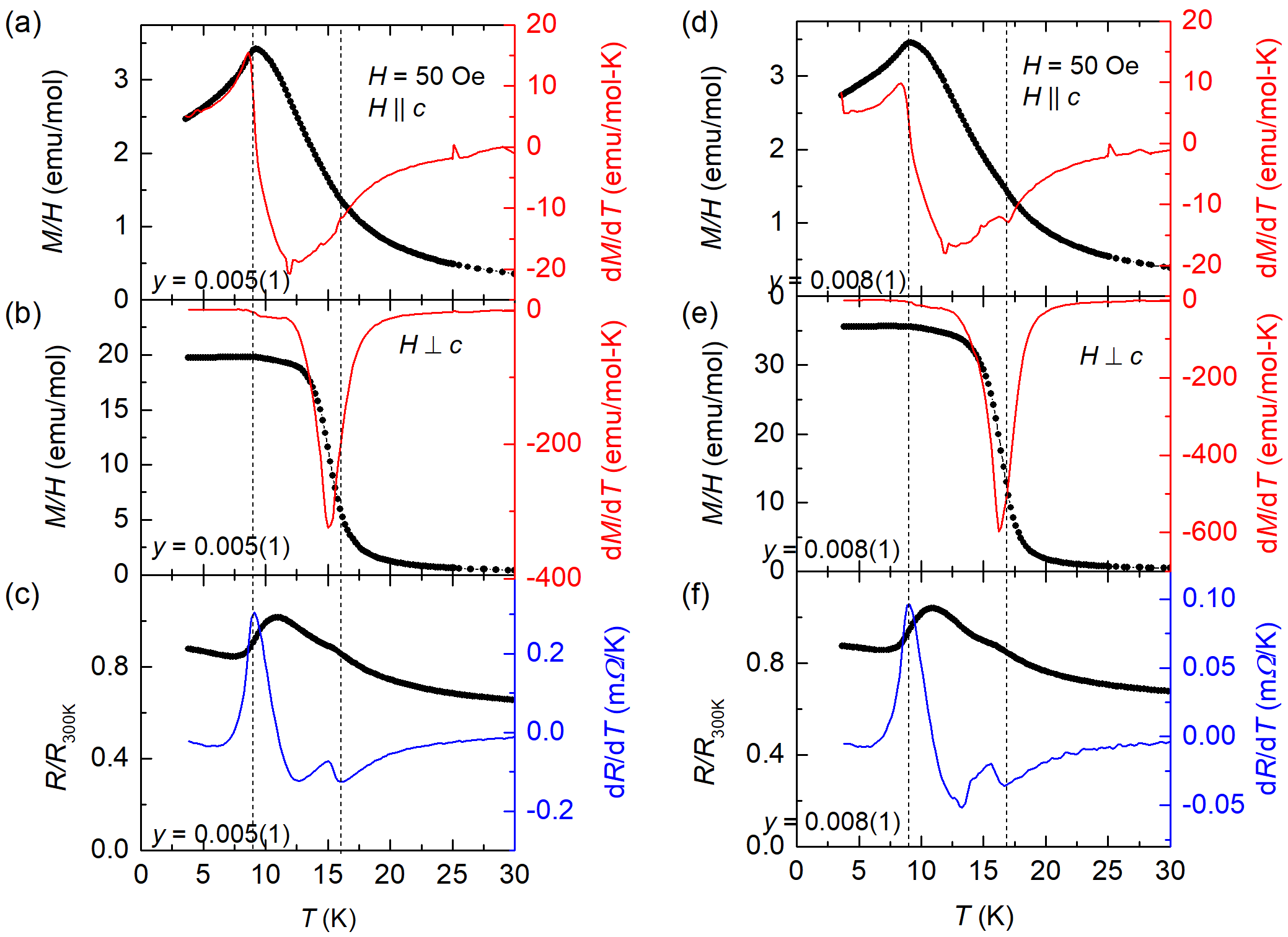}
	\caption{The low temperature dependences of normalized resistance $R/R_{300 K}$, the temperature derivative of the normalized resistance $d[R(T)/R_{300 K}]/dT$, temperature dependent magnetization divided by field $M(T)/H$, and the derivative $dM/dT$ plotted together for $T\leq30~$K for $y=0.005(1)$ and $x=0.008(1)$ Na substituted samples. (a) $M(T)/H$ plotted along with the temperature derivative $dM/dT$ for $H = 50~$Oe field cooled in $H \parallel c$ direction and (b) $H \perp c$ direction for $y=0.005(1)$ substitution. (c) $R/R_{300 K}$ and $d[R(T)/R_{300 K}]/dT$ for $x=0.005(1)$ substitution. The peak positions in $d[R(T)/R_{300 K}]/dT$, more or less coincides with the feature in $M(T)/H$. Panels (d), (e), and (f) show similar data for $x=0.008(1)$ sample. The peak positions in $d[R(T)/R_{300 K}]/dT$, more or less coincides with the feature in $M(T)/H$. The low field $M(T)$ and zero field $R(T)$ have small Sn features, therefore we truncate data below $T_c$ of Sn. Dotted lines show the transition temperatures determined from $d[R(T)/R_{300 K}]/dT$.}
	\label{mtrt2}
\end{figure*}

We will first present the temperature dependent electrical resistance and magnetization data, examine the effects of Na- and Ag-substitutions on the ground state as well as the higher temperature properties and construct a temperature-substitution phase diagram. Figure \ref{rt} (a) and (b) shows the resistivity $\rho$ as a function of temperature for Ag and Na-substituted samples on a semi-log scale, with the inset showing low temperature behavior for the substituted samples more clearly. Figure \ref{rhoArr} in the appendix shows the data for the undoped EuCd$_2$As$_2$ in an Arrhenius plot, showing that below roughly $50~$K the temperature dependence of the resistivity data can be fit well with an $\sim$ 5-6 meV gap. Although the absolute value of resistivity for the undoped sample might be prone to error because of the relatively high contact resistances, the orders of magnitude difference is probably inherent to the sample itself. It is also worth noting that this value differs from that reported in the literature, \cite{Ma2019, Du2022} but given that EuCd$_2$As$_2$ appears to be close to the realm of semiconducting physics, where small doping changes the resistivity immensely, widely varying resistivity depending on the differing sample quality are not surprising. High purity Ames Laboratory Eu, along with other elements used in the synthesis of samples for this study may lead to lower accidental doping levels, thereby changing the resistivity drastically. As soon as we start chemical substitution the resistivity values drop two orders of magnitude at $300~$K, and the significant upturn in resistance of undoped EuCd$_2$As$_2$ became less pronounced, but remains clearly resolvable (Fig. \ref{rt} inset). The upturn in the undoped sample has been attributed to strong magnetic fluctuations existing in this low carrier density system, \cite{Rahn2018, Ma2019} or formation of ferromagnetic clusters in similar systems. \cite{Sunko2022} Hence a reduced up turn might be denoting reduced magnetic fluctuations, possibly along with a higher carrier density leading to more metallic-like behavior. Two clear features (denoted by red arrows for $x=0.0055(3)$ sample as an example) are seen in the low temperature resistance data for both Ag and Na doped samples, shown in the inset of Figs. \ref{rt} (a) and (b), corresponding to the two transitions which will be discussed in detail later in the text. 

Temperature dependent magnetization, $M(T)$, divided by applied field, $H$, for $3.8~$K$\leq T\leq35~$K, for various Ag-substituted samples are shown in Fig. \ref{mt_Ag} (a) for applied magnetic field parallel to the crystallographic $c$-axis and (b) for field in the $ab-$plane. The data were taken while cooling down with the applied field of $50~$Oe (zero field cooled data shown in Appendix Fig. \ref{zfcfc}). There is roughly an order of magnitude difference between the $M(T)/H$ for the $H \parallel c$ and $H \perp c$ directions with the $c-$axis being the harder direction and the basal plane being the easier direction. Additionally, whereas the in-plane magnetization shows a rapid increase and lower temperature plateau like a transition with a ferromagnetic component, in the $T$ range 12-18$~$K for various samples, field parallel to $c-$axis data shows more complicated behavior. There is a clear transition around 10$~$K, below which $M/H$ is decreasing for samples with low substitution, and increasing for samples with high Ag-substitution. At the same time, there also exists a less noticeable feature at a higher temperature (e. g. $T\sim18~$K for $x=0.004$ sample), which shifts further upward with increased doping. 

Figure \ref{mt_Na} shows the temperature dependent magnetization (field cooled) data for the Na doped samples. The behavior of $M(T)/H$ for the Na-substituted samples is very similar to that of Ag-substituted samples. Crystallographic $c-$ axis remains the harder direction with $M(T)/H$ values with applied field along the $c-$axis being almost 10 times smaller than the in-plane magnetization. The in-plane $M(T)/H$ shows a sudden jump around $15-17~$K for both the Na doped samples, whereas magnetization with applied field parallel to $c-$axis shows a clear feature around $10~$K with a subtle higher temperature ($\sim17~$K) feature for the $y=0.008$ sample, similar to the Ag doped samples.  

In order to follow the features in $R/R_{300 K}(T)$ and $M(T)/H$ more clearly and to understand their behavior, we have plotted the magnetization divided by field $M(T)/H$ measured at $H=50~$Oe in both directions with it's temperature derivatives $dM/dT$ along with the normalized resistance $R(T)/R_{300~K}$ measured at zero field, and derivative of resistance $d[R(T)/R_{300~K}]/dT$, for two samples with Ag-substitution in Fig. \ref{mtrt1} and Na substituted samples in Fig. \ref{mtrt2}. Similar plots for the rest of the Ag doped samples are shown in Appendix Fig. \ref{mtrt3}.

\begin{figure*}
	\includegraphics[scale=1]{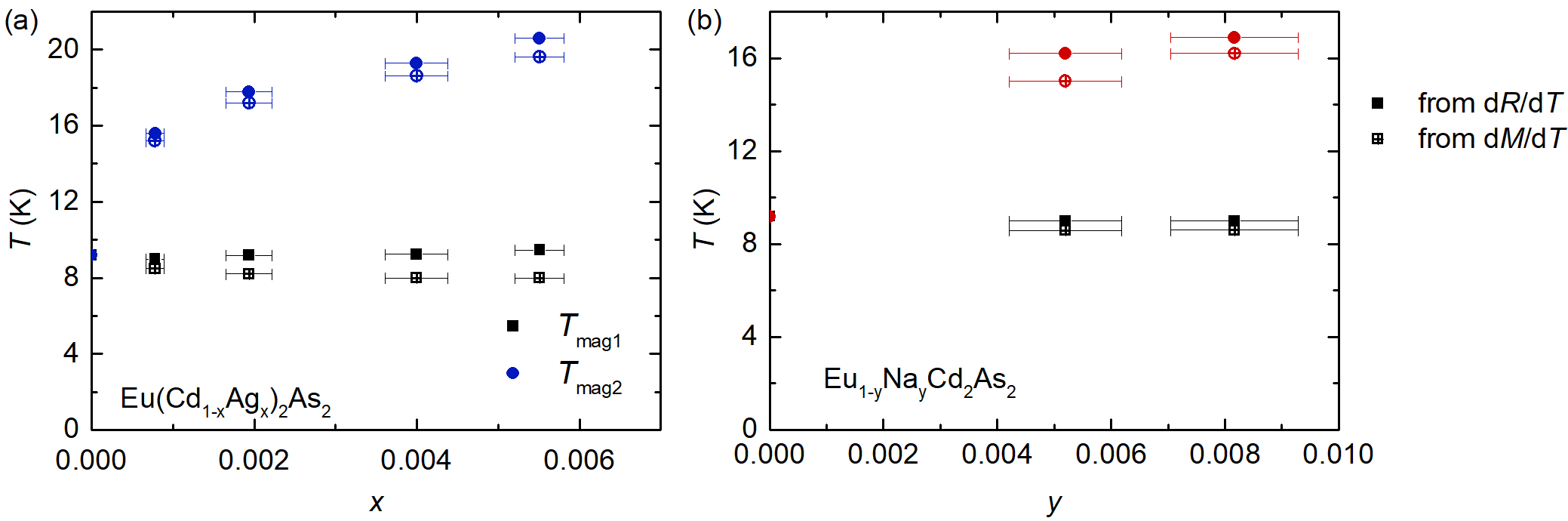}
	\caption{(a) Temperature-doping ($T-x$) phase diagrams for Ag substituted EuCd$_2$As$_2$. Open-crossed and closed symbols show the transition temperatures from $dM/dT$ and $d[R(T)/R_{300 K}]/dT$ respectively. (b) Temperature-doping ($T-y$) phase diagrams for Na substituted EuCd$_2$As$_2$  
	}
	\label{pd1}
\end{figure*}

Figures \ref{mtrt1} (a) and (b) show the anisotropic $M(T)/H$ and $dM/dT$ for the $x=0.0008$ Ag substituted sample. Figure \ref{mtrt1} (c) shows the normalized resistance and temperature derivative of the normalized resistance for the same sample. Similar plots for Ag doping $x=0.0040$ are shown in Figs. \ref{mtrt1} (d), (e), and (f). Whereas the $R(T)/R_{300~K}$ data, and its temperature derivative show two clear features, the anisotropic $M(T)/H$ sets each show only one conspicuous feature. For $H \perp c$, the rapid increase in magnetization that suggests a ferromagnetic component to the ordering agrees well with the higher temperature feature in the resistance data. For $H\parallel c$, the clearer feature in the $M(T)/H$ data agrees with the lower temperature feature seen in the $R(T)/R_{300~K}$ data, and its temperature derivative. For higher Ag-substitution levels there may be a small higher temperature shoulder associated with the higher transition temperature for $H\parallel c$, but this is subtle. The two transitions from $d[R(T)/R_{300~K}]/dT$ and $dM/dT$ have been identified as $T_{mag1}$ and $T_{mag2}$, and are plotted in the phase diagram in Fig. \ref{pd1} (a).     

A similar analysis has been carried out on the data for the Na-substituted compound. Fig. \ref{mtrt2} (a) and (b) show the anisotropic $M(T)/H$ and $dM/dT$ for the $y=0.005$ Na substituted sample. Figure \ref{mtrt2} (c) shows the normalized resistance and temperature derivative of the normalized resistance for the same sample. Similar plots for Na doping $y=0.008$ are shown in Figs. \ref{mtrt2} (d), (e), and (f). The two features are clearly seen in the resistance data and its derivatives shown in blue. Anisotropic $M(T)/H$ shows features similar to Ag-substituted samples. Whereas for $H\perp c$, $M(T)/H$ shows a sudden increase, with the transition temperature agreeing with the higher temperature feature in $R(T)/R_{300~K}$, $H\parallel c$ data has a prominent feature corresponding to the lower temperature transition and a subtle feature for the higher temperature one. The transition temperatures $T_{mag1}$ and $T_{mag2}$ determined form $d[R(T)/R_{300~K}]/dT$ and $dM/dT$ are used to plot the phase diagram in Fig. \ref{pd1} (b).  

The transition temperatures obtained from $R(T)$ and $M(T)$ curves can now be used to construct $T-x$ phase diagrams as shown in Fig. \ref{pd1} for Ag- and Na-substituted EuCd$_2$As$_2$. The transition temperatures were determined by taking the extrema in $d[R/R_{300~K}]/dT$ \cite{Fisher1968} and $dM/dT$. For Ag- and Na-substitution (Fig. \ref{pd1} (a) and (b) respectively), one can clearly see $T_{mag1}$ remaining more or less constant with substitution, and with finite, and increasing substitution a new transition, at $T_{mag2}$, appears and increases with substitution level. We can try to understand this along with the insights from the magnetization and resistance data as follows: With Ag- and Na-substitution in EuCd$_2$As$_2$, we are splitting the magnetic transition that was initially at $T_N=9.2~$K in to two, one that is denoted by $T_{mag1}$ that is remaining unchanged, and one denoted by $T_{mag2}$ that shows a gradual increase with increased chemical substitution. One can further speculate, the transition corresponding to $T_{mag2}$ is associated with a ferromagnetic alignment of moments in the $ab-$plane, as would be supported by the prominent feature in the in-plane $M/H$ measurement. In this case, the chemical substitution seems to play a key role in splitting a transition into two separate transitions, which is similar to what was observed in La doped EuPd$_3$S$_4$. \cite{Ryan2022} 

\begin{figure*}
	\includegraphics[width=\linewidth]{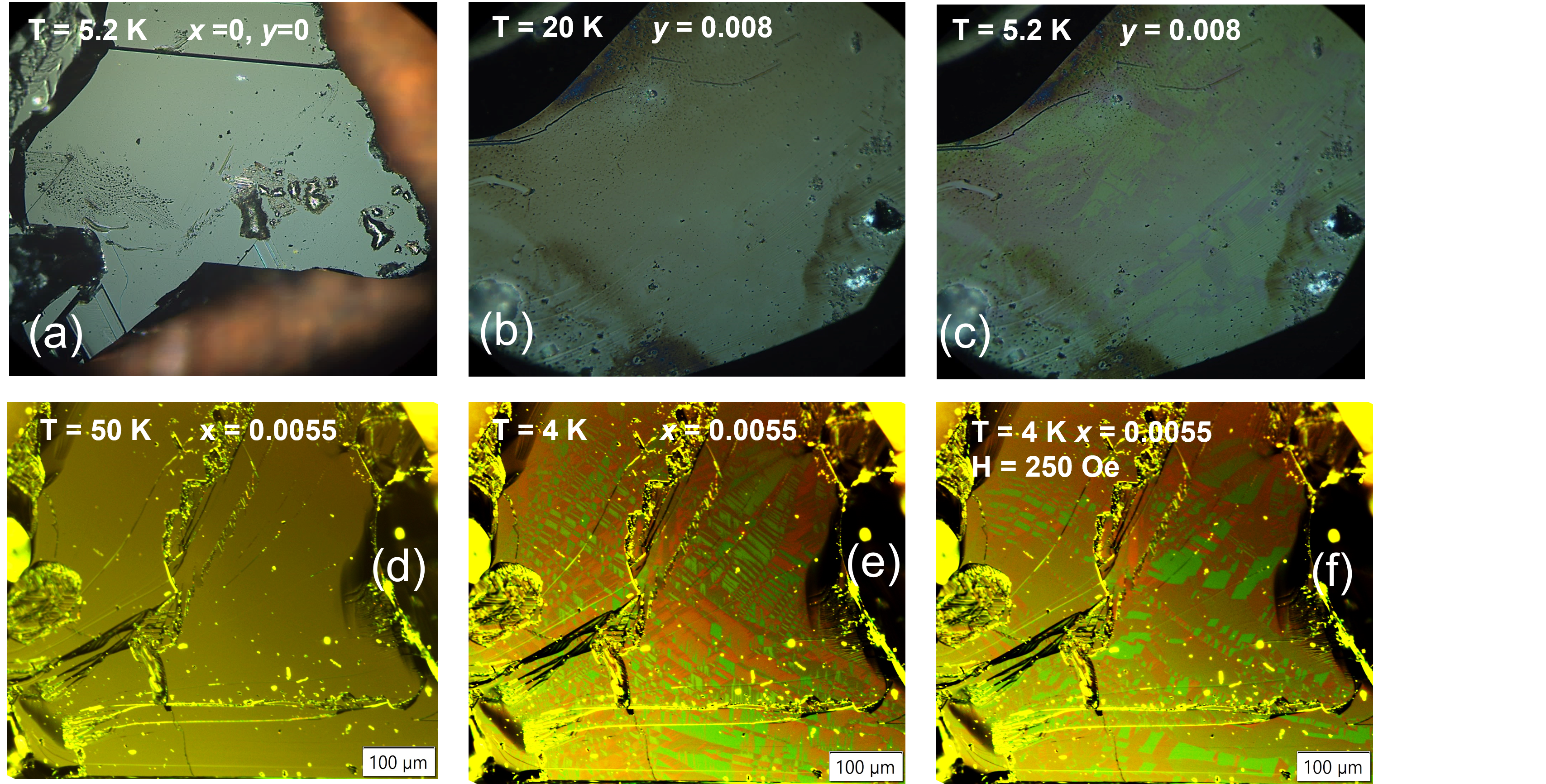}
	\caption{Magneto optical images of EuCd$_2$As$_2$, Ag doped sample $x = 0.0055$, and  Na doped sample with $y = 0.008$. (a) Undoped EuCd$_2$As$_2$ at $T=5.2~$K, which is below $T_N$, not showing magnetic domains. (b) Eu$_{0.992}$Na$_{0.008}$Cd$_2$As$_2$ crystal at $T=20~$K, above $T_{mag2}$, not showing any contrast due to domain formation. (c) Same sample at $T=5.2~$K, below both magnetic transitions, after cooling in zero field and $H=0$, showing ferromagnetic domain formation seen as green and purple colored regions in the images. Similar images for Eu(Cd$_{0.9945}$Ag$_{0.0055}$)$_2$As$_2$ shown in (d) at $T=50~$K above T$_{mag2}$ and (e) at $T=4~$K, with no applied field. (f) Eu(Cd$_{0.9945}$Ag$_{0.0055}$)$_2$As$_2$ sample at $T=4~$K with an in-plane magnetic field $H=250~$Oe applied. Both (e) and (f) show formation of domains.}
	\label{mo}
\end{figure*}

\subsection{Magneto-optical Kerr effect}

Before studying the field dependent magnetization and magnetoresistance of these samples, it is useful to further confirm that the low temperature state of these substituted samples indeed have a zero field, ferromagnetic component. As was the case for the EuCd$_2$As$_2$ samples grown from NaCl/KCl salt mixtures \cite{Jo2020} magneto optical measurements, shown in Fig. \ref{mo}, show the formation of clear magnetic domains below the transition temperatures for $x = 0.0055$ and $y = 0.008$ Ag- and Na-substitution. Figure \ref{mo} (a) shows the pure EuCd$_2$As$_2$ sample, where no domains are visible even below the magnetic transition temperature. Furthermore, Figs. \ref{mo} (b) and (d) show the Na and Ag doped samples $y=0.008$ and $x=0.0055$, at $20~$K and $50~$K respectively, above both magnetic transitions. Once again, these show no domain formation. At lower temperatures, both these samples show clear domain formation as shown in Figs. \ref{mo} (c) and (e). Comparison of Figs. \ref{mo} (e) and (f) demonstrates that the application of $250~$Oe along the $ab-$plane changes the size of the observed domains, which indicates that the domains are indeed magnetic in origin. Similar experiments were conducted with application of a magnetic field along the $c-$axis. Even though this is the harder magnetic axis, domains showed a change in size and shape (Appendix Fig. \ref{mo_sm}). These data demonstrate that at low temperatures, the zero field magnetic state in the doped samples has an unambiguous ferromagnetic component in the $ab-$plane.  

\begin{figure*}
	\includegraphics[scale=1]{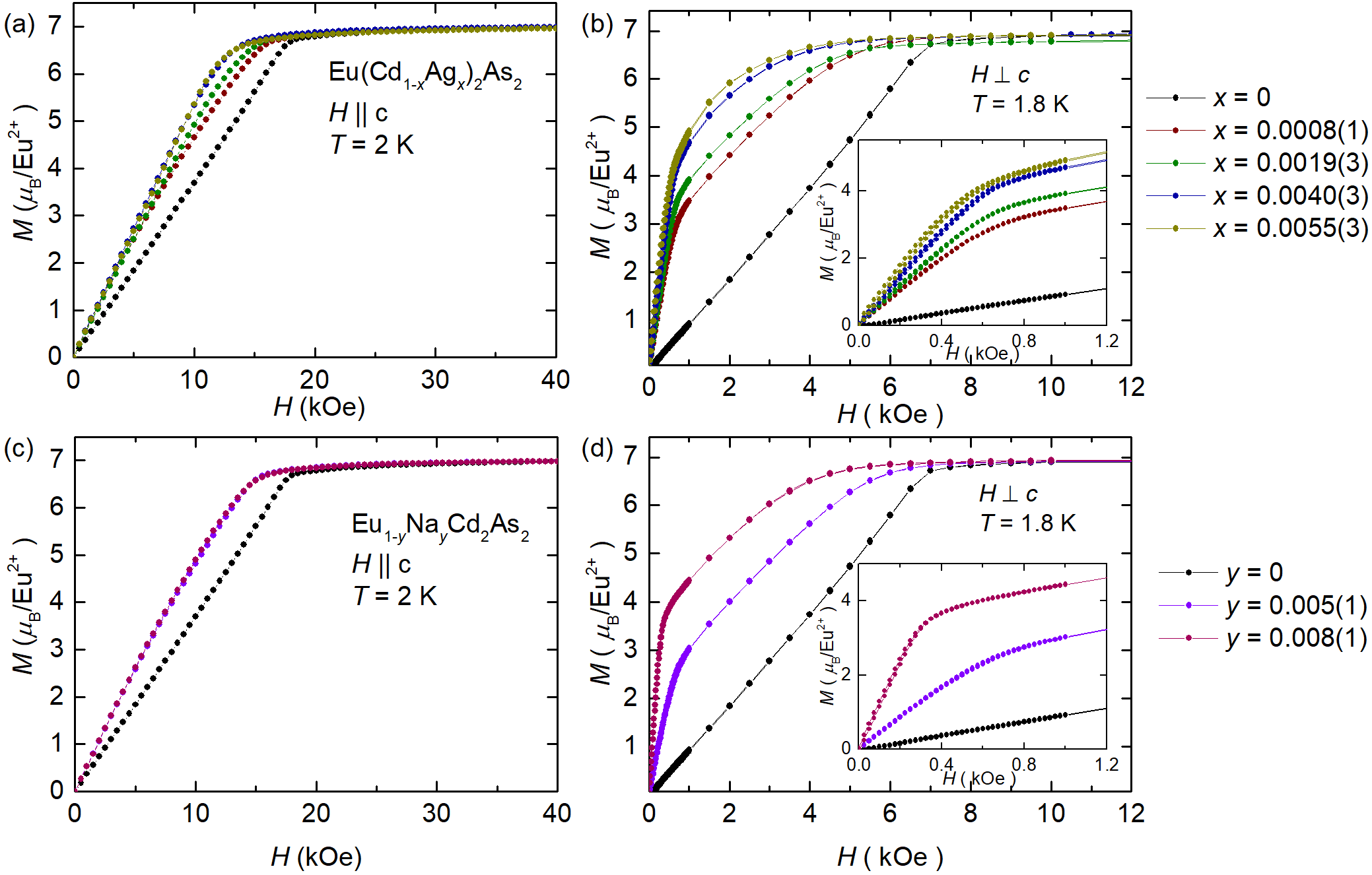}
	\caption{(a) Magnetization $M$ as a function of $H$, with applied field $H\parallel c$, for Eu(Cd$_{1-x}$Ag$_x$)$_2$As$_2$ samples. (b) $M(H)$ with field $H\perp c$ for Ag doped EuCd$_2$AS$_2$ samples. Inset: $M(H)$ at the low field region $H<1.2~$kOe for $H\perp c$ direction, showing the sharp rise. (c) Magnetization $M$ as a function of $H$, with applied field $H\parallel c$, for Eu$_{1-y}$Na$_y$Cd$_2$As$_2$ samples. (d) $M(H)$ with field $H\perp c$ for Na doped EuCd$_2$AS$_2$ samples. Inset: $M(H)$ at the low field region $H<1.2~$kOe. All $M(H)$ data were normalized to give the saturated moment of $7~\mu_B$, which is the value expected for Eu$^{2+}$.}
	\label{mh}
\end{figure*}
\begin{figure*}
	\includegraphics[scale=1]{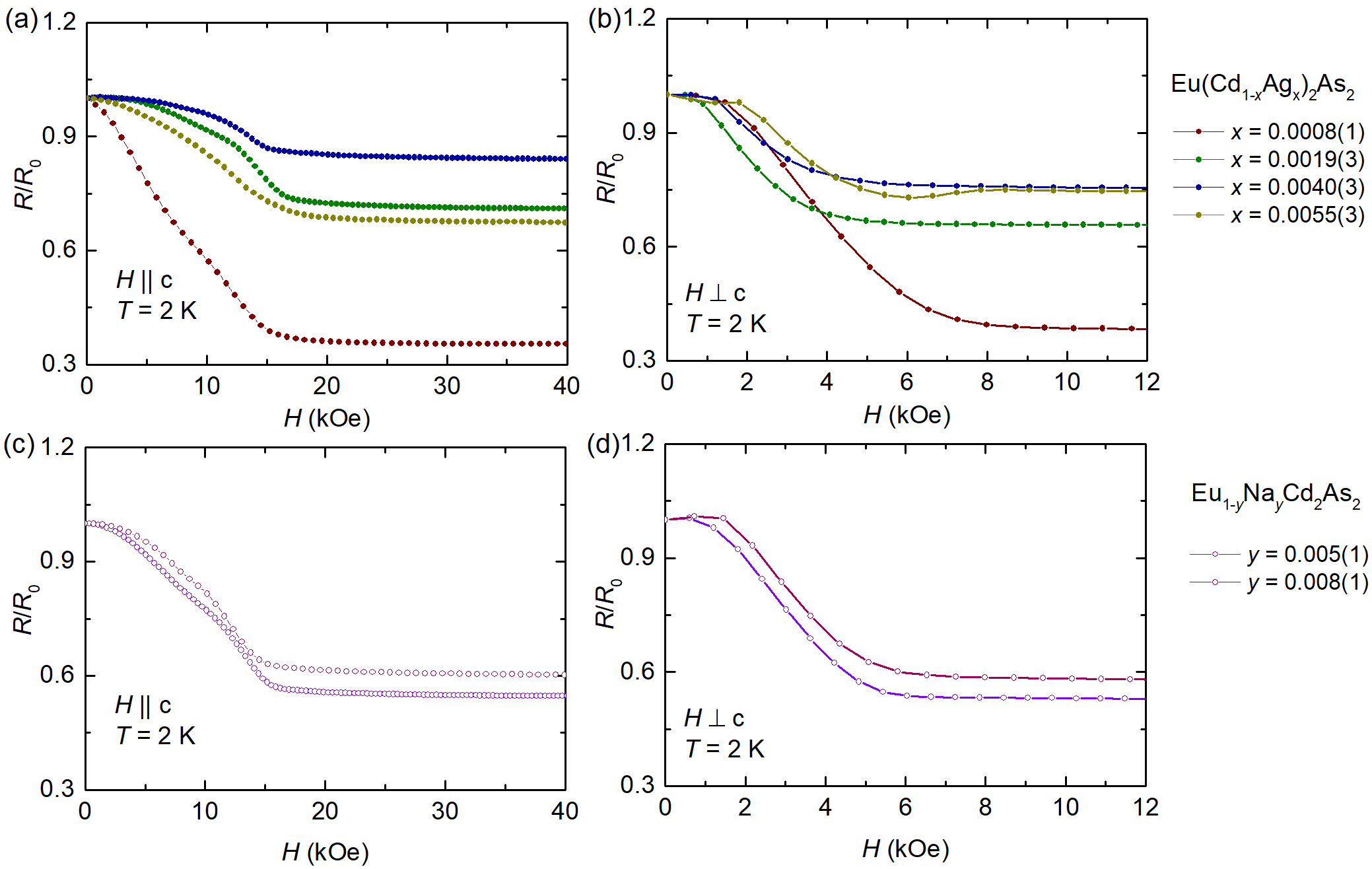}
	\caption{Normalized resistance $R/R_0$ (where $R_0$ is the resistance at zero field) as a function of $H$ for Ag-substituted samples with (a) $H\parallel c$ and (b) $H \perp c$. Panels (c) and (d) show the measurement results for Na-substituted samples}
	\label{rh}
\end{figure*}

\subsection{Magnetization and magnetoresistance}
\begin{figure*}
	\includegraphics[width=\linewidth]{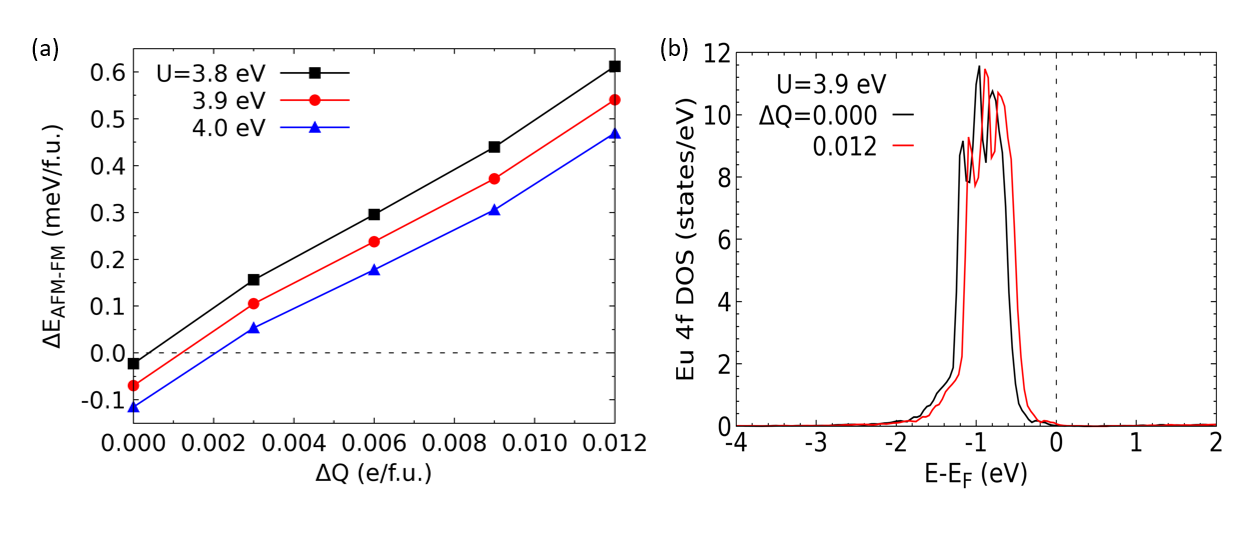}
	\caption{(a) Energy difference $\Delta E_{(AFM-FM)}$ in meV per formula unit (meV/f. u.) as a function of doping $(\Delta Q)$ for various values of $U$. Positive values in $\Delta E$ show the doping region for which FM state is stable. (b)  The density of states (DOS) projected on Eu 4f, showing the change of the distance between Eu 4f band and $E_F$ upon doping.}
	\label{dft}
\end{figure*}
In order to gain further insight into how Ag- and Na-substitutions affect the magnetic ground state of EuCd$_2$As$_2$ we have performed anisotropic $M(H)$ and $R(H)$ measurements on our samples. The base temperature anisotropic $M(H)$ for Ag-substituted EuCd$_2$As$_2$ data is shown in Fig. \ref{mh} (a) and (b) whereas the anisotropic $M(H)$ for Na-substituted EuCd$_2$As$_2$ data is shown in Fig. \ref{mh} (c) and (d). All $M(H)$ data were normalized to give the saturated moment of $7~\mu_B$, which is the value expected for Eu$^{2+}$. This is to avoid sample dependence and errors due to different mounting. Figure \ref{rh} shows the base temperature anisotropic $R(H)$ data for Ag- and Na-substituted EuCd$_2$As$_2$. In the appendix (Figs. \ref{Ag5} - \ref{Na20}) we plot the $M(H)$ (without normalization) and $R(H)$ data for each sample and direction of field separately for more direct comparison of field induced features. 

From the $M(H)$ data we can see that as Ag and Na are substituted, a ferromagnetic component develops in the basal plane, whereas the changes in the $M(H)$ behavior for $H\parallel c$ are far more subtle. For $H \perp c$, as $x$ or $y$ increases there is a dramatic increase in the low field slope and, as shown in the previous section, the establishment of a ferromagnetic component along with ferromagnetic domains. There is a step-like change in the magnetization between pure EuCd$_2$As$_2$ and the lowest $x-$ or $y-$values with the pure sample being in a clear antiferromagnetic state and the samples with finite Ag- or Na-substitution exhibiting $M(H)$ data consistent with a degree of ferromagnetic ordering. The rapid rise in $M(H)$ persists to roughly $0.5~$kOe, after which the $M(H)$ enters an intermediate state, finally saturating by $\sim 8~$kOe. As $x$ or $y$ increases the size of the moment associated with the low field roll over increases, reaching roughly $4~\mu_B$ per Eu$^{2+}$ (out of the $7~\mu_B$ associated with fully saturated Eu$^{2+}$). The temperature evolution of the various features in $M(H)$ for the Ag doped sample with $x=0.0019$, with applied field $H\perp c$ was studied systematically, and plotted in the $H-T$ phase diagram shown in Fig. \ref{htpd} in Appendix. 

For $H\parallel c$ the effect of substitution is relatively small. Whereas for pure EuCd$_2$As$_2$ it takes $\sim 17~$kOe to complete a spin flip process, Ag- and Na-substitution lower this field to $\sim 12$ or $\sim 15~$kOe respectively. As will be discussed below, there are subtle changes in the slope of $M(H)$ below this saturation field, but these changes are more clearly inferred from the $R(H)$ data.

The anisotropic $R(H)$ data is shown in Fig. \ref{rh}. Figure \ref{rh} (a) and (b) show the anisotropic $R(H)$ data for the Ag substituted samples and Fig. \ref{rh} (c) and (d) show the data for Na substituted samples. The primary point that can be extracted from Fig. \ref{rh} is that, for each direction of applied field, the $R(H)$ saturates once the $M(H)$ saturates. Given that magnetic anisotropy seen in Fig. \ref{mh}, the saturation in $R(H)$ occurs around $5-8~$kOe for $H \perp c$ and around $15-18~$kOe for $H\parallel c$. It is worth emphasizing that the $M(H)$ and $R(H)$ data demonstrate that for pure and doped EuCd$_2$As$_2$, the base temperature $R(H)$ is dominated by scattering off of the Eu$^{2+}$ moment and its excitations.
 
More detailed insight can be drawn from individual comparison of the anisotropic $M(H)$ and $R(H)$ data shown for each sample in Figs. \ref{Ag5}-\ref{Na20} in the appendix section. For $H \perp c$, the majority of the change in resistance occurs between the $\sim 0.5~$kOe saturation of the FM component and the $\sim5-8~$kOe saturation. This suggests that whereas aligning the FM domains does not significantly change scattering, aligning the Eu moments through what is most likely some form of spin rotation leads to a significant reduction of scattering. Once the Eu moments are fully aligned with the applied field the $R(H)$ saturates and becomes essentially field independent. For $H\parallel c$, $R(H)$ data show clear signs of an intermediate field transition. This is particularly clear for the Na-doped samples (Figs. \ref{Na10} and \ref{Na20} ). For example, for $y=0.005$, although the $M(H)$ and $R(H)$ data saturate around $17~$kOe, there is a broad feature in $R(H)$ around $6~$kOe. At the same field the $M(H)$ has a subtle, but discernible, change to a slightly shallower, higher field, slope. Similar features in both $H\parallel c$, $R(H)$ and $M(H)$ can be seen for other Na-substituted sample as well as some of the Ag-substituted ones. These features suggest that for $H\parallel c$ there may be an intermediate field change in ordered state. Ultimately this will need to be examined by diffraction measurements.

Given the ability to compare $R(H)$ and $M(H)$ for the Ag- and Na-substituted samples it is worthwhile to compare these data with $R(H)$ data taken under pressure. \cite{Gati2021} In Ref. \onlinecite{Gati2021} the saturation field for the $H \perp c$, $R(H)$ data was tracked as a function of pressure and when it was suppressed to zero a transition to a pressure induced FM state was inferred. As we can now see for Ag- and Na-substituted samples, we can have a FM component to our zero field ground state without having reached a saturation point in the $R(H)$ data. Indeed, what we do see is that for the Ag- and Na- substituted samples we have studied the $R(H)$ is decreasing with field rather than increasing (as it does for pure EuCd$_2$As$_2$ at ambient pressure as well as for $p \lesssim 1.67~$GPa) when the sample is in a state with an FM component. Based on the Ag- and Na-substituted data we may now speculate that for $1.67~$GPa $<p<2.05$~GPa, pure EuCd$_2$As$_2$ may be in a magnetic state with a ferromagnetic component similar to what we find for Ag- and Na- substituted compounds. This range of pressure is also where an insulator to metal transition at low temperatures is reported for the undoped compound. \cite{Du2022} For $p > 2.05~$GPa pure EuCd$_2$As$_2$ would then be in a FM state with full Eu moments all co-aligned (i.e. fully saturated). This possibility was already allowed for in Ref. \onlinecite{Gati2021}, specifically in the shading/color coding of the temperature - pressure phase diagram shown in its figure 3 and associated text.

\subsection{DFT Calculations}

Since chemical substitution both at the magnetic Eu site (by Na), and at the non magnetic Cd site (by Ag), lead to similar magnetic behavior and $T-x$ phase diagrams, one likely mechanism behind it can be a change in band filling/density of states near the Fermi energy ($E_F$) and its effect on magnetic coupling. This is further supported by the DFT calculation shown in Fig. \ref{dft}. In rare-earth containing compounds, the magnetic interaction between the $f-$electrons is often mediated by the conduction electrons through the $RKKY$ mechanism. In the case of EuCd$_2$As$_2$, Eu $4f$ magnetic exchange interaction between layers is mediated by the free electron like hole pockets from As $p$ derived bands around $E_F$. The preference of AFM or FM ground state is affected by the energy separation ($\sim 1.0~$eV) between Eu $4f$ and As $p$ bands at $E_F$, which can be tuned by exchange splitting strength $U$ and number of electrons removed $\Delta Q$. The calculated difference in energy between the AFM-A state and FM state, $\Delta E(AFM-FM)(\Delta Q)$ for various values of $U$ are shown in Fig. \ref{dft} (a), capturing the FM phase being stabilized by very small changes in band filling at $E_F$, consistent with the substitution levels actually used in this work. The calculations were made with the moments along one of the equivalent $a-$ or $b-$axis. The $U$ values chosen are close to those in recent studies,\cite{Gati2021, Yu2022} which use lower values of $U$ than earlier studies, \cite{Wang2019, Jo2020} in order to predict the transitions with pressure more precisely. Fig. \ref{dft} (b) shows the density of state (DOS) projected on Eu 4f, showing the change of the distance between Eu 4f band and $E_F$ upon doping. When electrons being removed, i. e. hole doping, moves Eu-$4f-$ DOS closer to $E_F$, the inter layer FM coupling becomes more favorable than AFM.  

\section{Conclusion}
In summary, we have synthesized single crystals of Eu(Cd$_{1-x}$Ag$_x$)$_2$As$_2$ and Eu$_{1-y}$Na$_y$Cd$_2$As$_2$ and constructed temperature-substitution phase diagrams. We have shown that Ag and Na substituted EuCd$_2$As$_2$ shows a splitting of a single magnetic transition in pure EuCd$_2$As$_2$ into two transitions, with a ferromagnetic component stabilized below both the transitions. We also propose a change in band filling as the possible mechanism to stabilize ferromagnetism, supported by theoretical calculations. These data underscore just how delicately balanced the magnetic state in EuCd$_2$As$_2$ is and illustrates how chemical substitution and changing the band filling can be the pathways to stabilize a ferromagnetic ground state, and thereby possibly tune the various topological features predicted in the electronic structure of this material. 

\begin{acknowledgements}
We thank Warren Straszheim for the EDS/WDS measurements. We also acknowledge D. H. Ryan, T. J. Slade, and N. H. Jo for the useful discussions. Work at the Ames National Laboratory was supported by the U.S. Department of Energy, Office of Science, Basic Energy Sciences, Materials Sciences and Engineering Division. The Ames National Laboratory is operated for the U.S. Department of Energy by Iowa State University under Contract No. DEAC0207CH11358. B. K.  was supported by the Center for the Advancement of Topological Semimetals, an Energy Frontier Research Center funded by the U.S. DOE, Office of Basic Energy Sciences. L.X. was supported, in part, by the W. M. Keck Foundation and the Gordon and Betty Moore Foundations EPiQS Initiative through Grant GBMF4411.
\end{acknowledgements}

\section*{Appendix}

\subsection{X-ray diffraction}
Lattice parameters obtained from the Rietveld refinement of powder X-ray diffraction data are given in Table \ref{xrd} and \ref{xrd2}. For Ag-substituted data, the values given are average of three runs on three different samples from each batch of sample. The goodness of fit $wR$ value given in the table are the worst among various fits for each Ag-substituted sample. The refinements (for both Ag and  Na) were carried out with fixing the fraction of all the elements to full occupancy, as allowing them to vary ended up with nonphysical results. The variation in lattice parameters is not resolvable and within the error bars. 
\begin{table*}
	\begin{tabular}{c c c c c}
		\hline\hline
		Eu(Cd$_{1-x}$Ag$_x$)$_2$As$_2$ & $x = 0.0008(1)$ & $x = 0.0019(3)$ & $x = 0.0040(3)$ & $x = 0.0055(3)$\\
		\hline
		$a(\AA)$& 4.4410(4) &4.4407(5) &4.440(1) &4.4404(4)\\
		$c(\AA)$& 7.3292(3) &7.3296(2) &7.327(1) &7.328(3)\\
		$wR$(\%)&17.62 &19.19 &17.72 &17.01\\
		\hline\hline
	\end{tabular}
	\caption{The average lattice parameters obtained from Rietveld refinements of Ag substituted EuCd$_2$As$_2$. The goodness of fit $wR$ values given are the worst among the various fits on each sample.}
	\label{xrd}
\end{table*}

\begin{table*}
	\begin{tabular}{c c c c c}
		\hline\hline
		Eu$_{1-y}$Na$_y$Cd$_2$As$_2$ & $y=0$ &$y = 0.005(1)$ & $y = 0.008(1)$ \\
		\hline
		$a(\AA)$&4.4408(5) &4.4412(3) &4.4415(3) \\
		$c(\AA)$&7.3289(4)  &7.3291(3) &7.3284(3) \\
		$wR$(\%)&14.49  &12.27 & 13.98 \\
		\hline\hline
	\end{tabular}
	\caption{The lattice parameters obtained from Rietveld refinement of pure and Na substituted EuCd$_2$As$_2$.}
	\label{xrd2}
\end{table*}
\subsection{Temperature dependent conductivity of pure EuCd$_2$As$_2$}
\begin{figure}
	\includegraphics[scale=1]{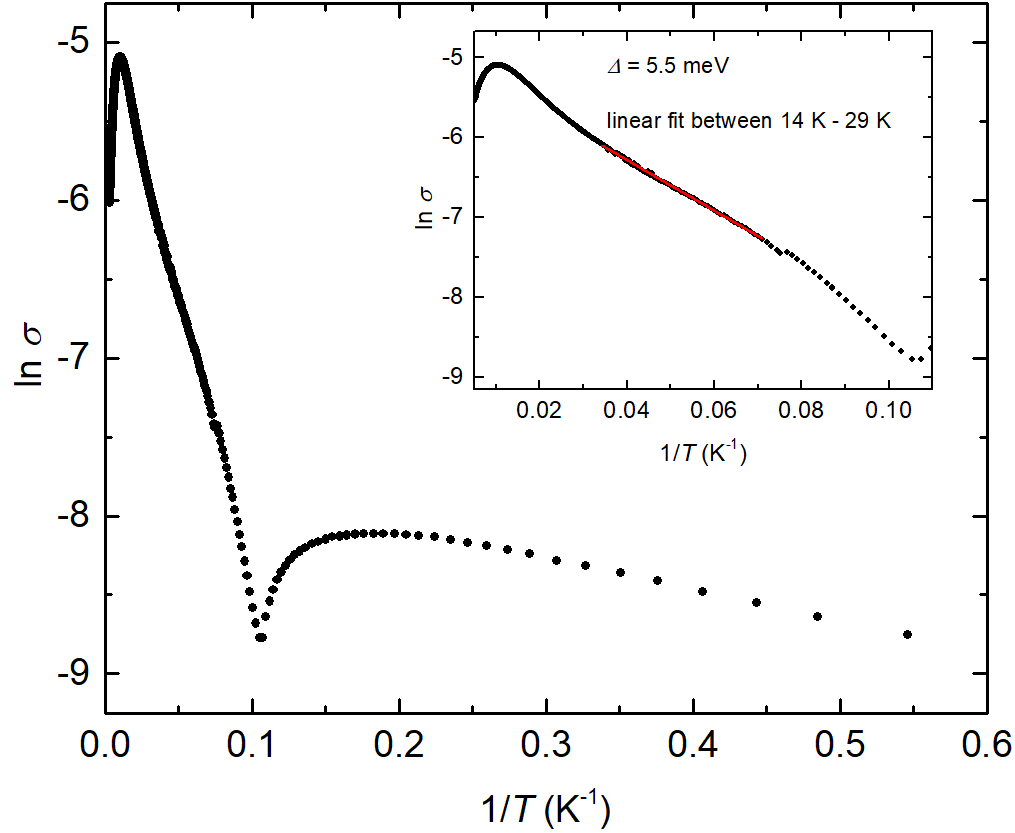}
	\caption{Arrhenius plot showing logarithm of electrical conductivity, $\ln\sigma$ as a function of $1/T$ for the undoped EuCd$_2$As$_2$. Inset shows the region where a linear fit has been done for temperatures between $14~$K and $29~$K. An energy gap of $\Delta =5.5~$meV is obtained from the slope, using $\sigma \propto \exp (-E_g/2k_BT)$.}
	\label{rhoArr}
\end{figure}

Figure \ref{rhoArr} presents the temperature dependent conductivity of EuCd$_2$As$_2$ on an Arrhenius plot. Below $\sim 50~$K there is an activated behavior the persists down to the magnetic ordering temperature, at which point there is a loss of spin-disorder scattering feature. The gap value that can be extracted from this behavior is roughly $5-6~$meV, consistent with the temperature range over which we see the Arrhenius behavior.

\subsection{Magnetization}

\begin{figure}
	\includegraphics[width=\columnwidth]{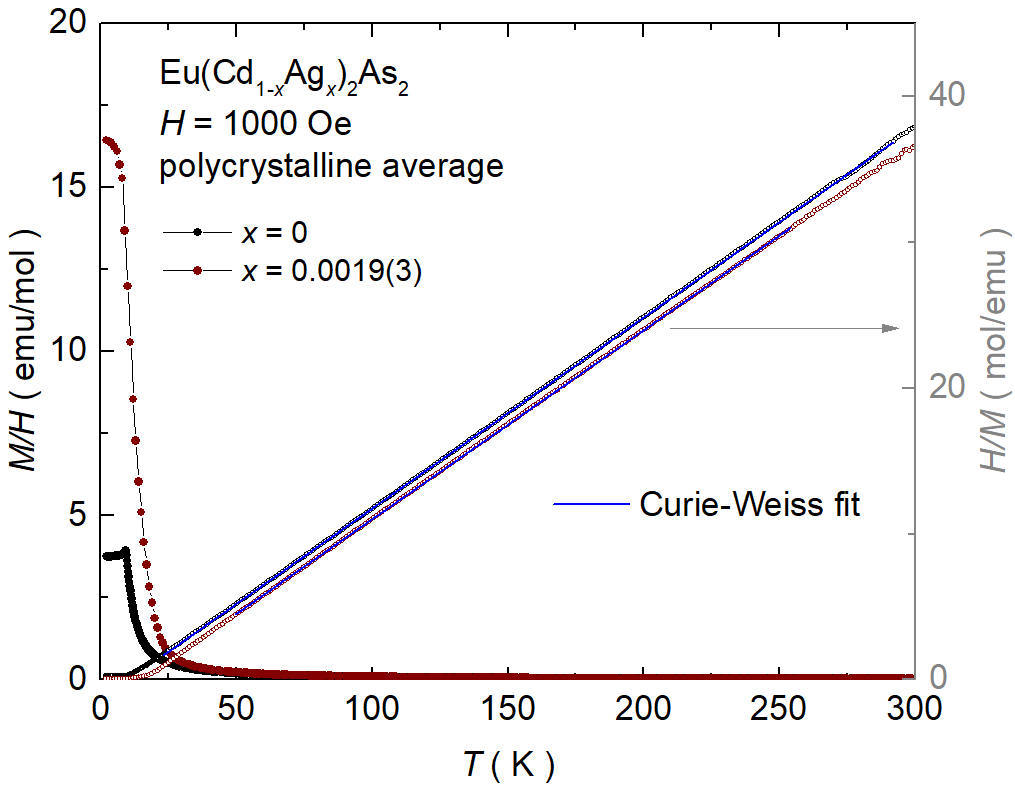}
	\caption{Polycrystalline average of magnetization divided by field $M/H$, as a function of temperature for undoped EuCd$_2$As$_2$, and Ag doping $x=0.0019$, with an applied field of $H=1~$kOe. Inverse susceptibility, $H/M$, plotted on the right axis for the two samples, along with Curie-Weiss fit for each. The effective moment and Weiss temperature obtained from the fitting parameters are given in Table \ref{cwt}.}
	\label{cw}
\end{figure}

\begin{figure}
	\includegraphics[scale=1]{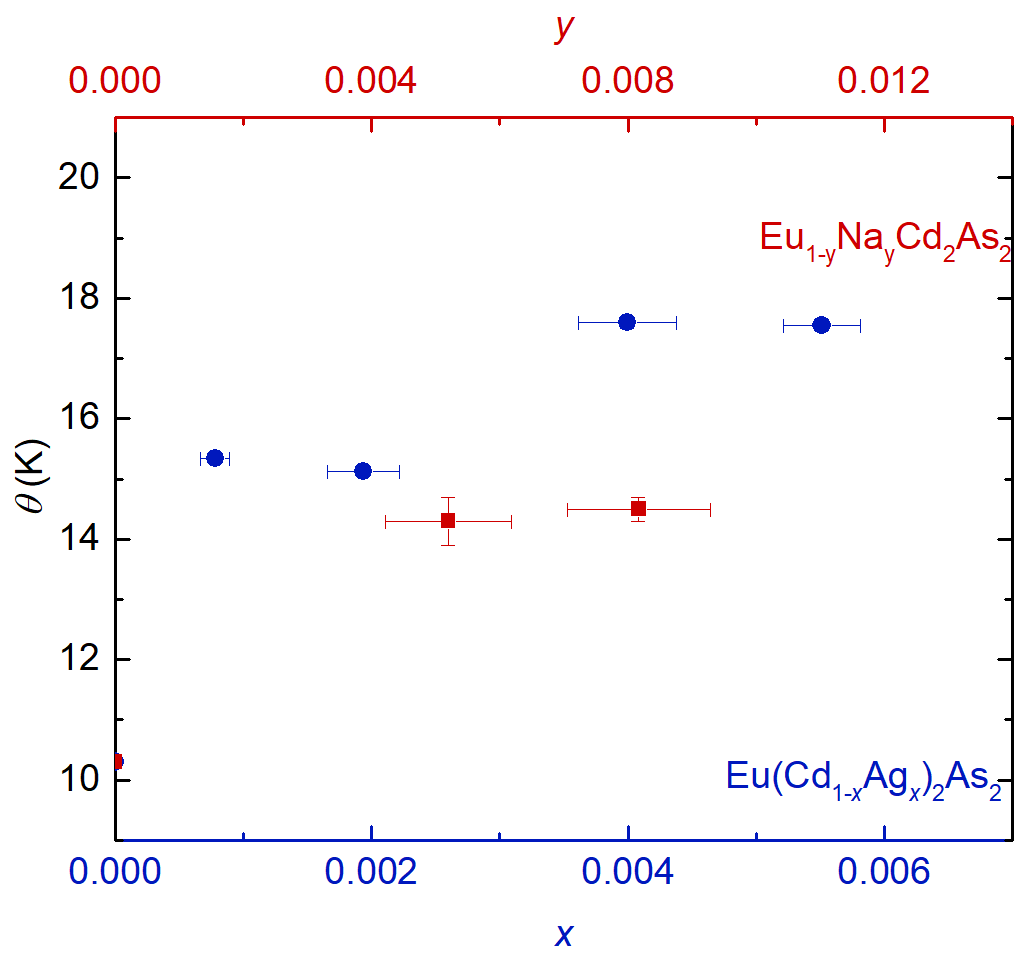}
	\caption{Curie-Weiss fitting parameter $\theta$ as a function of Ag- and Na-substitution, $x$ and $y$. Top axis has been scaled by two, as Na-substitution nominally changes 1 electron per formula unit, where as Ag-substitution changes two electrons per formula unit, as there are two Cd atoms.}
	\label{theta}
\end{figure}

The Curie-Weiss fit of the inverse of magnetic susceptibility, $H/M$, in the temperature range $50-250~$K, is shown in Fig. \ref{cw} for the undoped EuCd$_2$As$_2$ and Ag-doped sample with $x=0.0019$. The fitting parameters give an effective moment, $\mu_{eff}=7.82\pm0.02~\mu_B$, and Weiss temperature, $\theta=10.30\pm0.04$~K for the pure compound. Similar fits were done on the inverse of polycrystalline averaged $M(T)/H$ for all the substituted samples. The values obtained from those are tabulated in Table. \ref{cwt}. The $\mu_{eff}$ values are more or less consistent with the expected value of $7.94~\mu_B$. The Weiss temperature $\theta$ varies with doping, showing a slight increase with doping. A plot showing the variation of $\theta$ with the substitution concentration $x$ and $y$ is shown in Fig. \ref{theta}. The bottom $x-$axis show the Ag concentration, and the top $x-$axis show Na concentration, scaled by a factor of two. This is because Ag doping corresponds to 2 electrons being removed per formula unit, as Ag atoms are replacing two Cd atoms in each formula unit. There is an increase in $\theta$ with doping, roughly corresponding to the increase in transition temperatures $T_{mag2}$.   

\begin{table}
	\begin{tabular}{c|c|c}
		\hline\hline
		Sample & $\mu_{eff}$($\mu_{B}$) & $\theta$ (K)\\
		\hline
		$x=0$, $y=0$ & 7.82$\pm$0.02 & 10.30$\pm$0.04 \\
		$x = 0.0008(1)$& 7.83$\pm$0.02&15.34$\pm$0.04\\
		$x = 0.0019(3)$&7.85$\pm$0.03&15.12$\pm$0.06\\
		$x = 0.0040(3)$&7.62$\pm$0.04&17.6$\pm$0.1\\
		$x = 0.0055(3)$&7.68$\pm$0.03&17.55$\pm$0.06\\
		$y = 0.005(1)$&7.7$\pm$0.2&14.2$\pm$0.4\\
		$y = 0.008(1)$&7.78$\pm$0.09&14.5$\pm$0.2\\
		\hline\hline
	\end{tabular}
	\caption{The effective moment and Weiss temperature for various Ag and Na substituted EuCd$_2$As$_2$ samples, obtained from Curie-Weiss fits to the inverse magnetic susceptibility data above the transition temperature. The error bars are calculated considering error in mass of the crystals and fitting errors.}
	\label{cwt}
\end{table}

\begin{figure*}
	\includegraphics[scale=1]{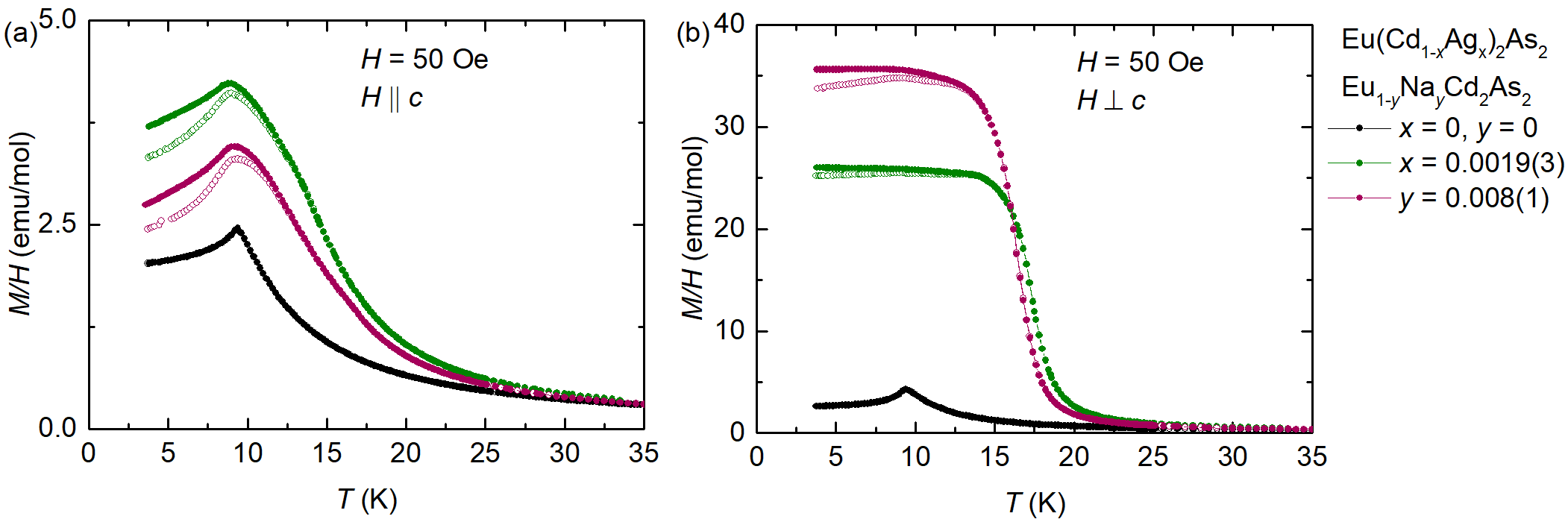}
	\caption{$M/H$ as a function of temperature for selected samples with an applied field of $H=50~$Oe. Measurements done with increasing temperature after zero field cooling (ZFC), and with decreasing temperature with field cooling (FC), denoted by open and closed symbols respectively. The data shown is above $T=3.8~$K, as there were slight features below that, due to presence of Sn (flux) in the sample which goes superconducting below this temperature. (a) Applied field parallel to crystallographic $c$ direction and (b) Field in the $ab-$plane.}
	\label{zfcfc}
\end{figure*}

Figure \ref{zfcfc} shows $M(T)/H$ data for zero field cooled (ZFC) and field cooled (FC) measurements for a selected set of Ag and Na substituted samples of EuCd$_2$As$_2$ above $T=3.8~$K to avoid features due to superconductivity from Sn flux present in some samples. The in-plane measurement shows almost overlapping data for ZFC and FC measurements on Ag substituted samples, whereas the ZFC and FC curves split with increased substitution with field $H\parallel c$. This could be indicating a possible non-collinear or helical order along the $c-$ direction with moments aligned ferromagnetically in-plane. The splitting of ZFC FC in both directions for the Na doped sample might be indicating a possible canting. 

\begin{figure*}
	\includegraphics[scale=0.8]{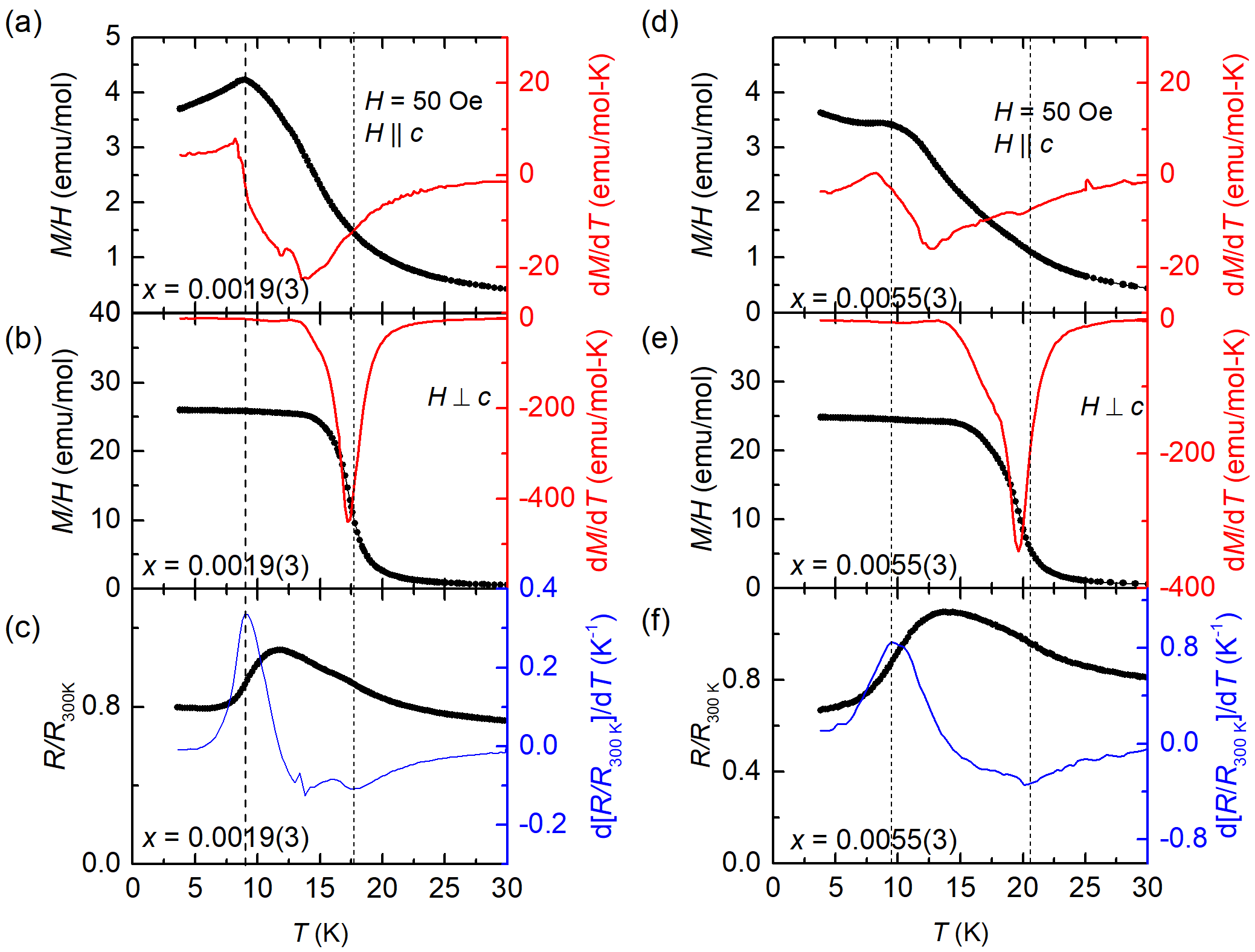}
	\caption{The low temperature dependences of normalized resistance $R/R_{300 K}$, the temperature derivative of the normalized resistance $d[R(T)/R_{300 K}]/dT$, temperature dependent magnetization divided by field $M(T)/H$, and the derivative $dM/dT$ plotted together for $T\leq30~$K for $y=0.005(1)$ and $x=0.008(1)$ Na substituted samples. (a) $M(T)/H$ plotted along with the temperature derivative $dM/dT$ for $H = 50~$Oe field cooled in $H \parallel c$ direction and (b) $H \perp c$ direction for $y=0.005(1)$ substitution. (c) $R/R_{300 K}$ and $d[R(T)/R_{300 K}]/dT$ for $x=0.005(1)$ substitution. The peak positions in $d[R(T)/R_{300 K}]/dT$, more or less coincides with the feature in $M(T)/H$. Panels (d), (e), and (f) show similar data for $x=0.008(1)$ sample. The peak positions in $d[R(T)/R_{300 K}]/dT$, more or less coincides with the feature in $M(T)/H$. The low field $M(T)$ and zero field $R(T)$ have small Sn features, therefore we truncate data below $T_c$ of Sn. Dotted lines show the transition temperatures determined from $d[R(T)/R_{300 K}]/dT$.}
	\label{mtrt3}
\end{figure*}

Figures \ref{mtrt3} (a) and (b) show the anisotropic $M(T)/H$ and $dM/dT$ for the $x=0.0019$ Ag substituted sample. Figure \ref{mtrt1} (c) shows the normalized resistance and temperature derivative of the normalized resistance for the same sample. Similar plots for Ag doping $x=0.0055$ are shown in Figs. \ref{mtrt3} (d), (e), and (f). Whereas the $R(T)/R_{300~K}$ data, and its temperature derivative show two clear features, the anisotropic $M(T)/H$ sets each show only one conspicuous feature. For $H \perp c$, the rapid increase in magnetization that suggests a ferromagnetic component to the ordering agrees well with the higher temperature feature in the resistance data. For $H\parallel c$, the clearer feature in the $M(T)/H$ data agrees with the lower temperature feature seen in the $R(T)/R_{300~K}$ data, and its temperature derivative.

\subsection{$H-T$ Phase diagram for Eu(Cd$_{1-x}$Ag$_x$)$_2$As$_2$ with $x=0.0019$ and applied field $H\perp c$}
\begin{figure*}
	\includegraphics[scale=1]{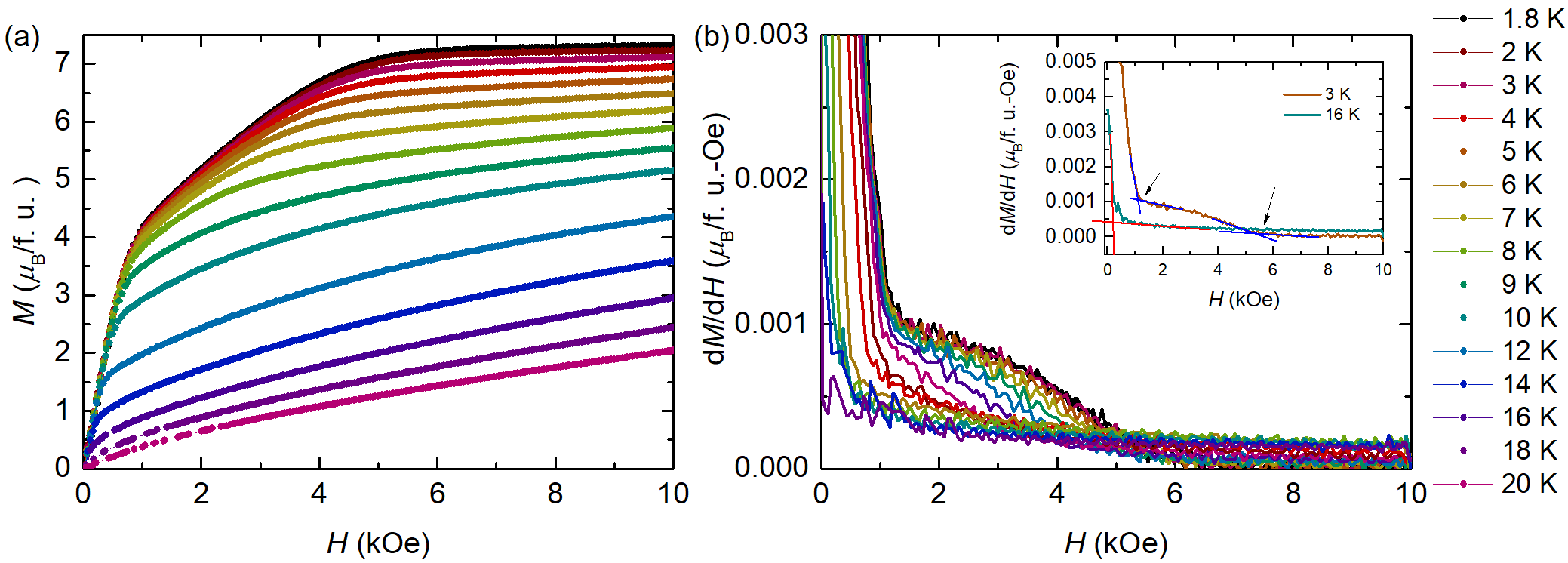}
	\caption{(a) Magnetization as a function of applied field, $M(H)$, up to $10~$kOe for various temperatures between $1.8~$K and $20~$K. (b) Derivative of magnetization with field, $dM/dH$ as a function of field  up to $10~$kOe for various temperatures between $1.8~$K and $20~$K. Inset shows the criteria for determining the transition points (marked by arrows pointing at them), where change of slope is determined by taking the intersection of straight lines drawn. All measurements shown are for field applied $H \perp c$, on the Ag doped sample with $x=0.0019$.}
	\label{MH_d}
\end{figure*}

\begin{figure*}
	\includegraphics[scale=1]{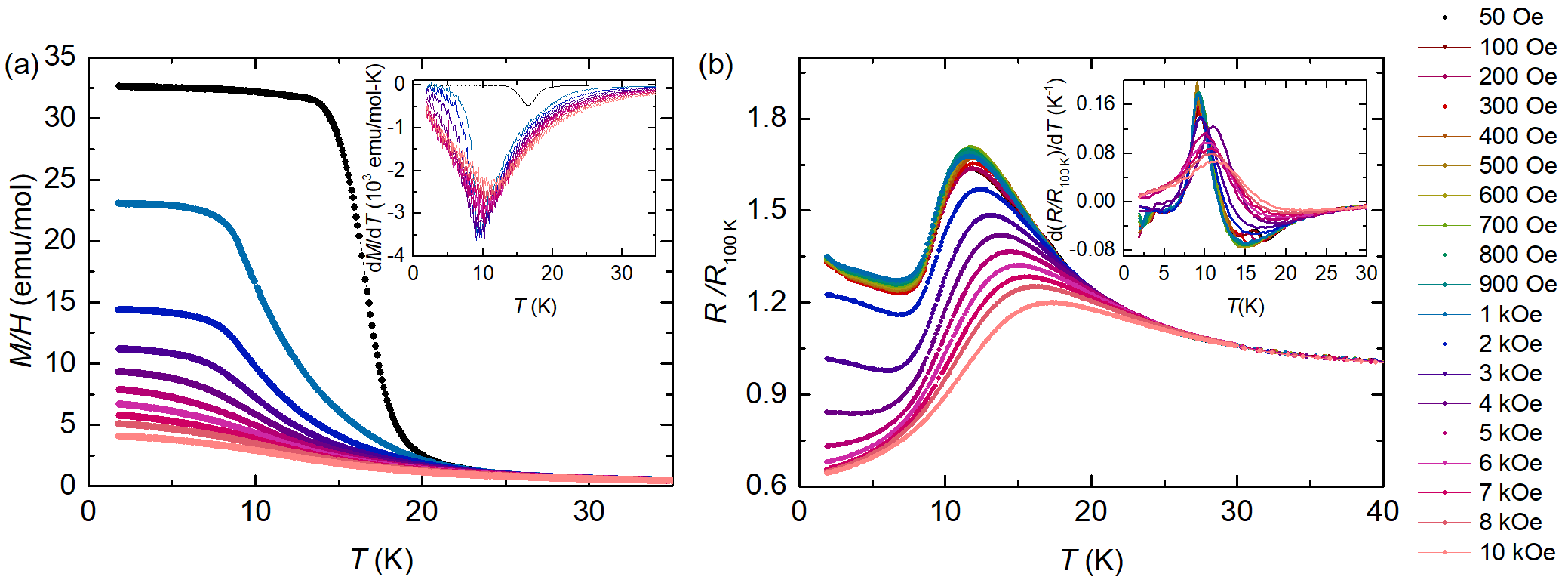}
	\caption{(a) Magnetization divided by field as a function of temperature, $M(T)/H$, for various applied fields between $50~$Oe and $10~$kOe. Inset: Derivative of magnetization with temperature, $dM/dT$, for various applied fields. The minima in these curves were chosen as the transition temperature for the $H-T$ phase diagram. (b) Normalized resistivity $R/R_{100~K}$ as a function of temperature  for various applied fields between $50~$Oe and $10~$kOe. Inset: Temperature derivative of normalized resistance, $d(R/R_{100~K})/dT$, for various applied fields. The extrema in these curves were chosen as the transition temperature for the $H-T$ phase diagram. All measurements shown are for field applied $H \perp c$, on the Ag doped sample with $x=0.0019$.}
	\label{MTRT_d}
\end{figure*}

\begin{figure}
	\includegraphics[scale=1]{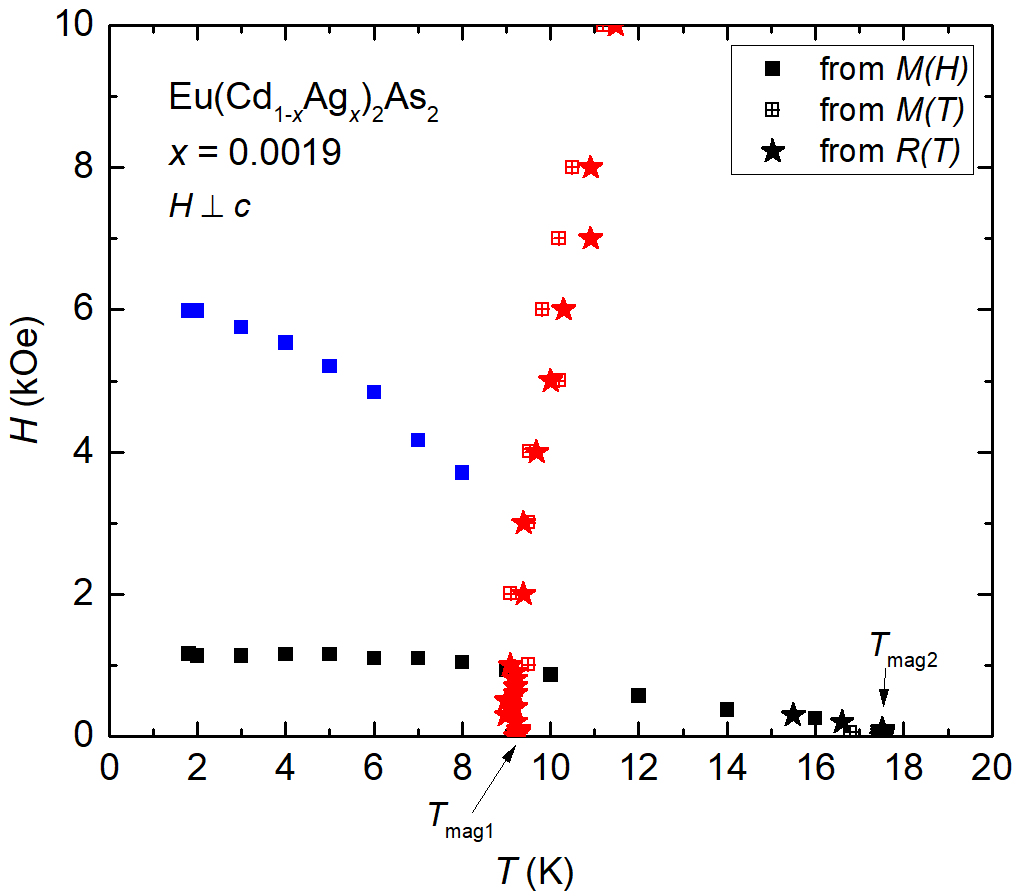}
	\caption{$H-T$ phase diagram obtained from various $M(T)$, $M(H)$, and $R(T)$ measurements on Eu(Cd$_{1-x}$Ag$_x$)$_2$As$_2$ with $x=0.0019$, for applied magnetic field $H \perp c$. Points obtained from $M(H)$ measurements at various temperatures are denoted by solid squares. Data points from $M(T)$ measurements at various applied fields are plotted in open-crossed symbols, and those from $R(T)$ at various fields are plotted with star symbols. Magnetic transition temperatures at zero field $T_{mag1}$ and $T_{mag2}$ are shown by arrows.}
	\label{htpd}
\end{figure}

The $T-x$ phase diagrams show existence of two magnetic phases with temperature in low fields. In addition to this, field dependent magnetization data $M(H)$ for Ag-doped samples with low doping, as well as Na-doped samples show, a clear intermediate field phase existing in these samples, with applied field $H \perp c$. In order see the temperature and field evolution of all these different magnetic phases, we did systematic measurements of $M(T)$ at various applied fields, $M(H)$ at various temperatures, and $R(T)$ at various applied fields, to construct an $H-T$ phase diagram for the $x=0.0019$ Ag doped EuCd$_2$As$_2$ sample. All the $M(T)$ and $R(T)$ data were measured on cooling the sample from above $T_{mag2}$, with various applied fields. $M(H)$ data were taken with increasing and decreasing magnetic fields up to $10~$kOe for each temperature. There was no discernible hysteresis, and the data shown in Fig. \ref{MH_d} were measured with decreasing fields. The transition temperatures and fields were determined from the derivatives, as shown in the figures below. 

Figure \ref{MH_d} (a) shows the $M(H)$ data taken at different temperatures, and the derivatives, $dM/dH$ are shown in Fig. \ref{MH_d} (b). The criteria for determining the fields corresponding to the transitions are shown in the inset. It can be seen that the intermediate - field phase gets narrower with increasing temperatures, and eventually disappears above $T_{mag1}$. The initial steep increase in $M$ with $H$, persists until higher temperatures, and disappears above $T_{mag2}$. This is consistent with the interpretation of initial increase due to alignment of domains, into a state with net ferromagnetic moment, and with higher fields all the moments getting aligned into a fully saturated state. The derivatives $dM/dH$ were plotted, and change of slope determined by drawing straight lines, to obtain the transition points. These points are plotted in Fig. \ref{htpd} using solid squares in black and blue.

Figure \ref{MTRT_d} (a) shows the $M(T)$ data at various applied fields between $50~$Oe to $10~$kOe. The inset shows the derivatives $dM/dT$, the minima of which was used to determine the transition temperatures. The data points obtained so are plotted as open-crossed symbols in the phase diagram in Fig. \ref{htpd}. Similarly $R(T)$ data taken at various applied fields are shown in Fig. \ref{MTRT_d} (b), with the inset showing the derivatives $dR/dT$. At low fields, the resistance curves and their derivatives show two transitions. With applied field more than $400~$Oe, the higher temperature transition is not resolvable, whereas the lower T transition remains independent of applied field at low and intermediate fields, and start moving to higher temperatures at larger fields. This behavior is traced by the extrema in $dR/dT$ curves, and plotted in Fig. \ref{htpd} with star shaped symbols, in black and red. 

The $H-T$ phase diagram obtained from the measurements mentioned above is plotted in Fig. \ref{htpd}. At lower field, there appears a dome like region below $\sim17~$K, which is presumably the phase with a net ferromagnetic moment, which shows domain formation and alignment. As we go to higher fields, for temperatures $T\leq T_{mag1}$, there appears another phase (whose boundary is denoted by the blue symbols), before eventually getting into a fully polarized state. There is also a nearly-vertical phase line, shown in red colored symbols, starting at $T_{mag1}$ at zero field. It remains mostly temperature independent with increasing fields, and crosses the black line around $H\sim 1~$kOe. With even higher fields, it seems to meet the blue line, and eventually moves towards slightly higher temperatures in this regime. This suggests one of the following possibilities allowed by the phase rules as a likely scenario to explain the phase diagram: 1. There is a first order structural transition (or electronic, Lifshitz-like) coinciding with the magnetic transition around $9~$K, which is then weakly magnetic field dependent, giving rise to the red line. But none of the previous studies have reported such a transition in the pure compound, which makes this scenario less likely. 2. The other possibility is that, we are falsely assigning a transition temperature and field in the Brillouin-like, slowly saturating with applied field, regime. In this case there is actually no phase transition happening along the red line but a slow crossover. The low field red points are likely the extension of the blue line. Thus the interpretation of this $H-T$ phase diagram remains an open question at this point.  

\subsection{Magnetization, Magnetoresistance and Magneto optics}
\begin{figure*}
	\includegraphics[scale=1]{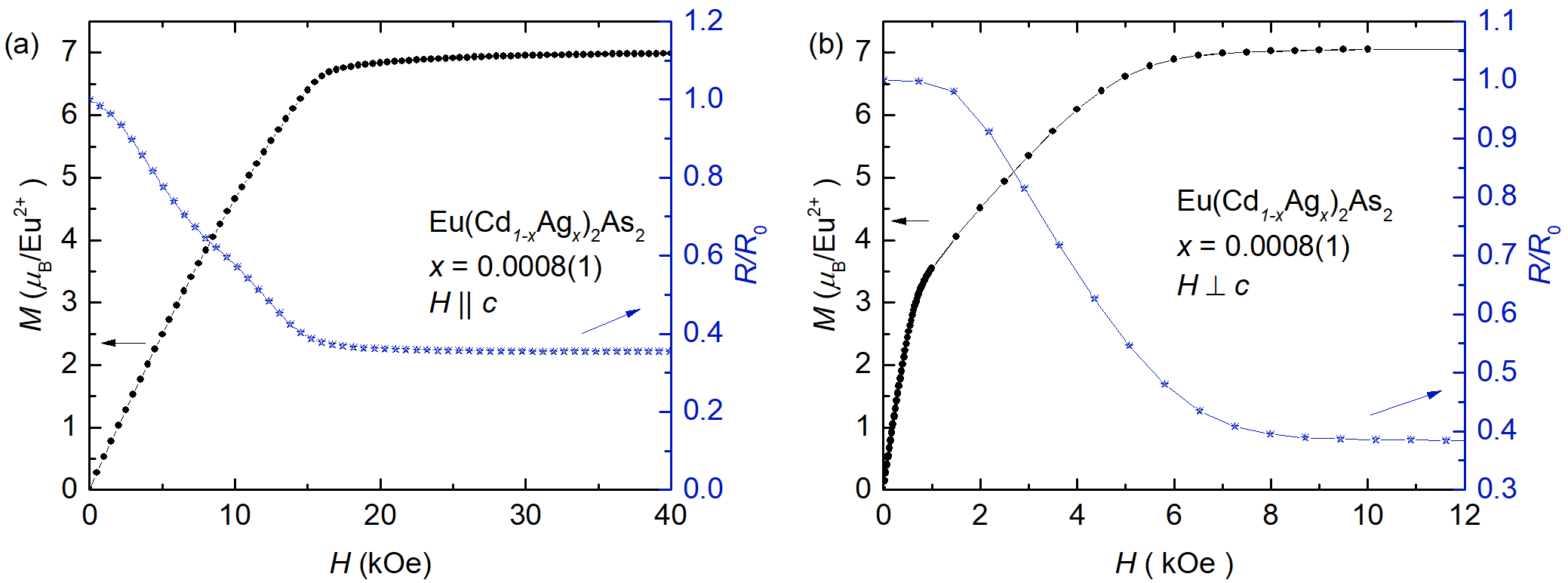}
	\caption{Magnetization as a function of field $M(H)$ measured at $T=1.8~$K and normalized resistivity as a function of field $R/R_0(H)$ measured at $T=2~$K for Ag doped EuCd$_2$As$_2$ with $x=0.0008(1)$. (a) Applied field $H\parallel c$ (b) $H \perp c$}
	\label{Ag5}
\end{figure*}

\begin{figure*}
	\includegraphics[scale=1]{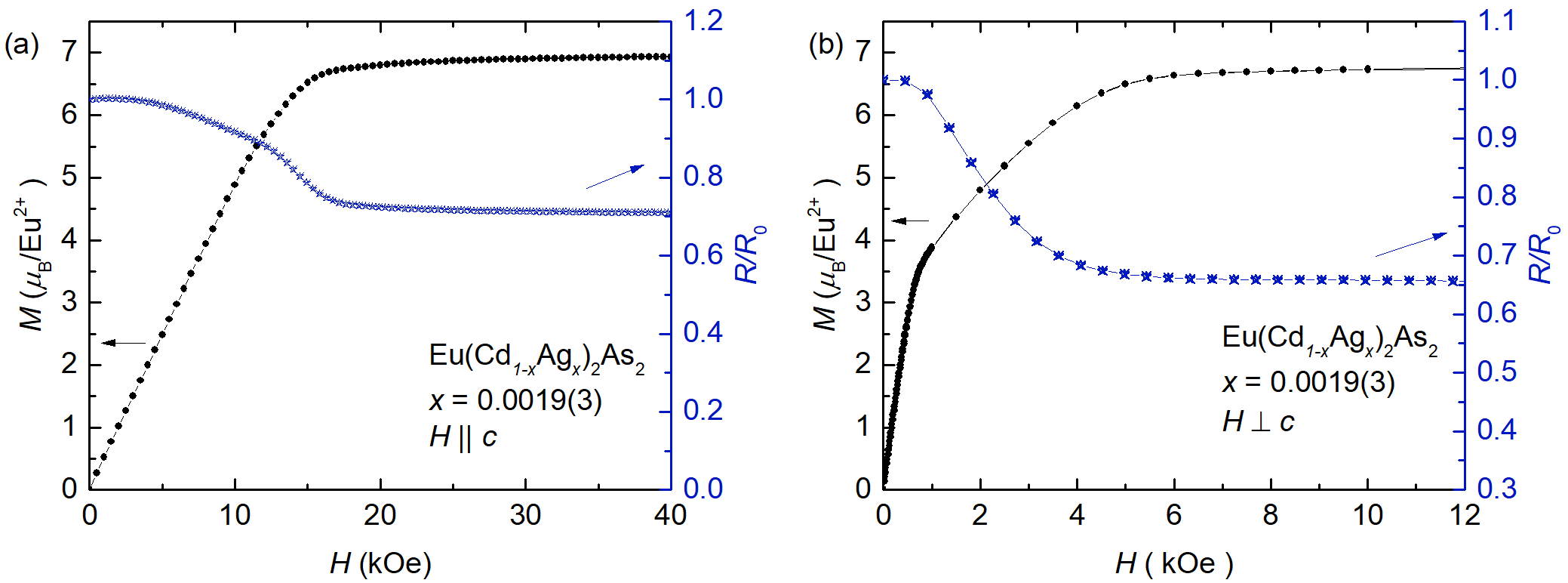}
	\caption{Magnetization as a function of field $M(H)$ measured at $T=1.8~$K and normalized resistivity as a function of field $R/R_0(H)$ measured at $T=2~$K for Ag doped EuCd$_2$As$_2$ with $x=0.0019(3)$. (a) Applied field $H\parallel c$ (b) $H \perp c$}
	\label{Ag10}
\end{figure*}

\begin{figure*}
	\includegraphics[scale=1]{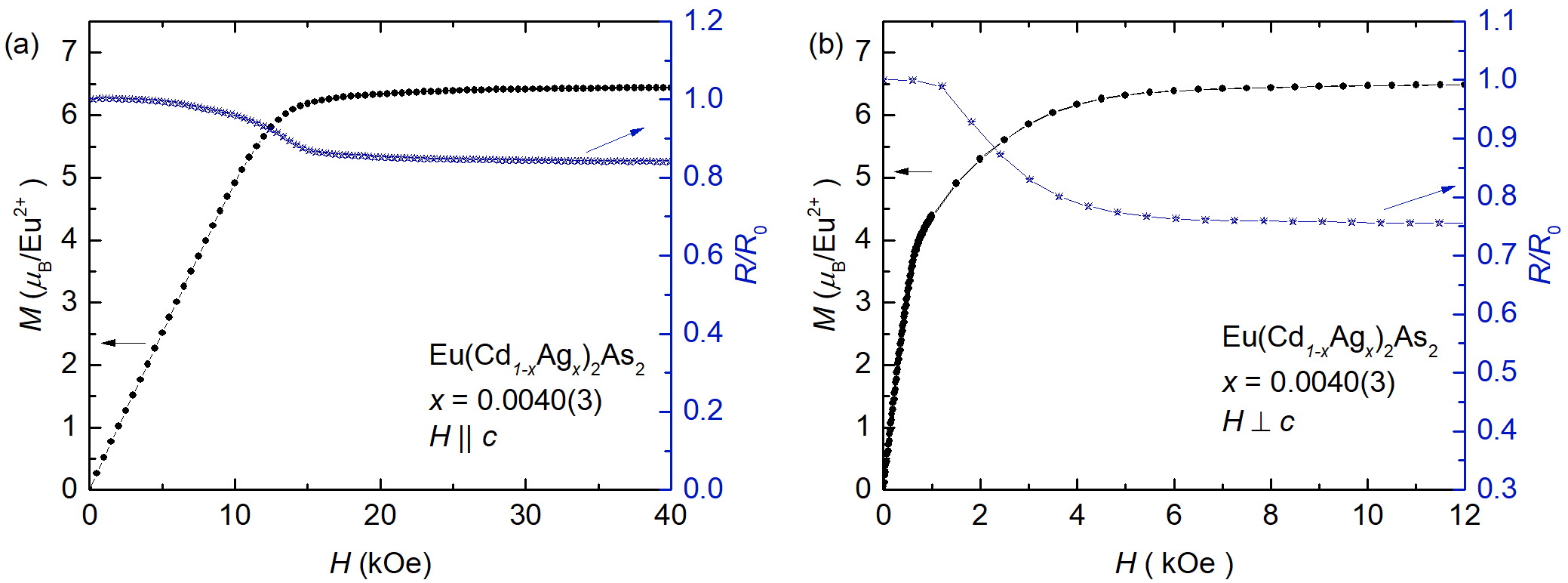}
	\caption{Magnetization as a function of field $M(H)$ measured at $T=1.8~$K and normalized resistivity as a function of field $R/R_0(H)$ measured at $T=2~$K for Ag doped EuCd$_2$As$_2$ with $x=0.0040(3)$. (a) Applied field $H\parallel c$ (b) $H \perp c$}
	\label{Ag20}
\end{figure*}

\begin{figure*}
	\includegraphics[scale=1]{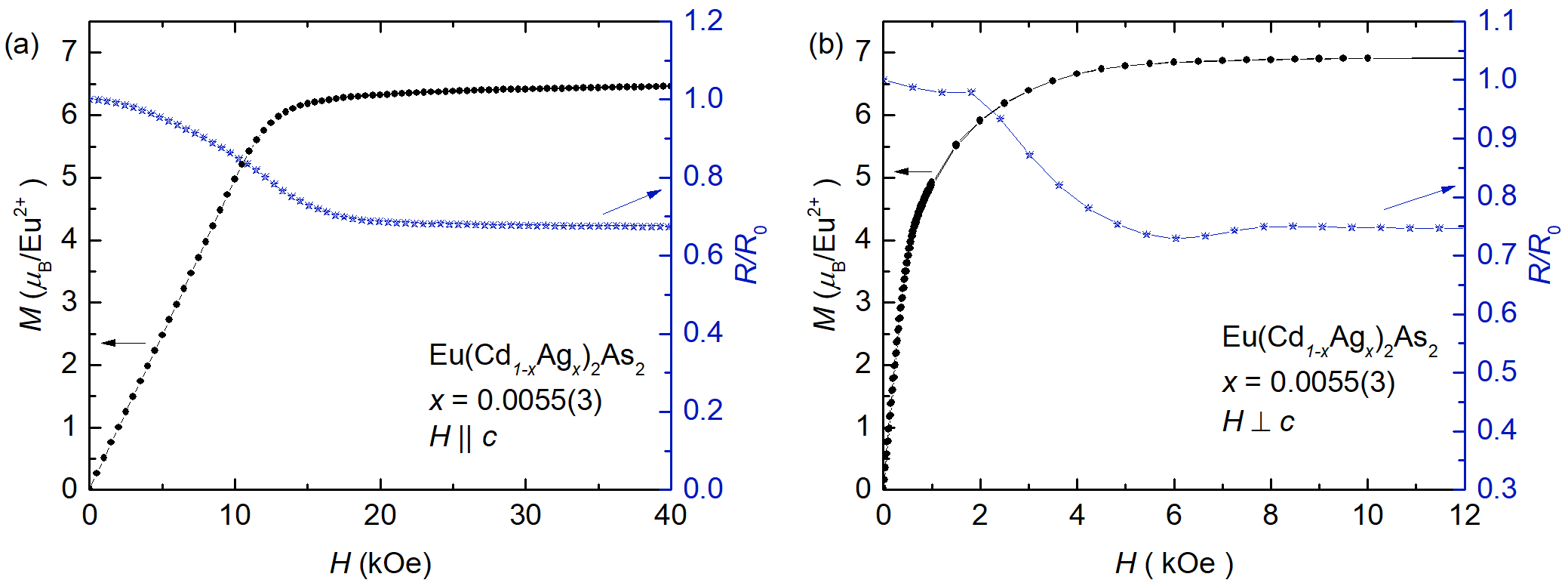}
	\caption{Magnetization as a function of field $M(H)$ measured at $T=1.8~$K and normalized resistivity as a function of field $R/R_0(H)$ measured at $T=2~$K for Ag doped EuCd$_2$As$_2$ with $x=0.0055(3)$. (a) Applied field $H\parallel c$ (b) $H \perp c$}
	\label{Ag30}
\end{figure*}

\begin{figure*}
	\includegraphics[scale=1]{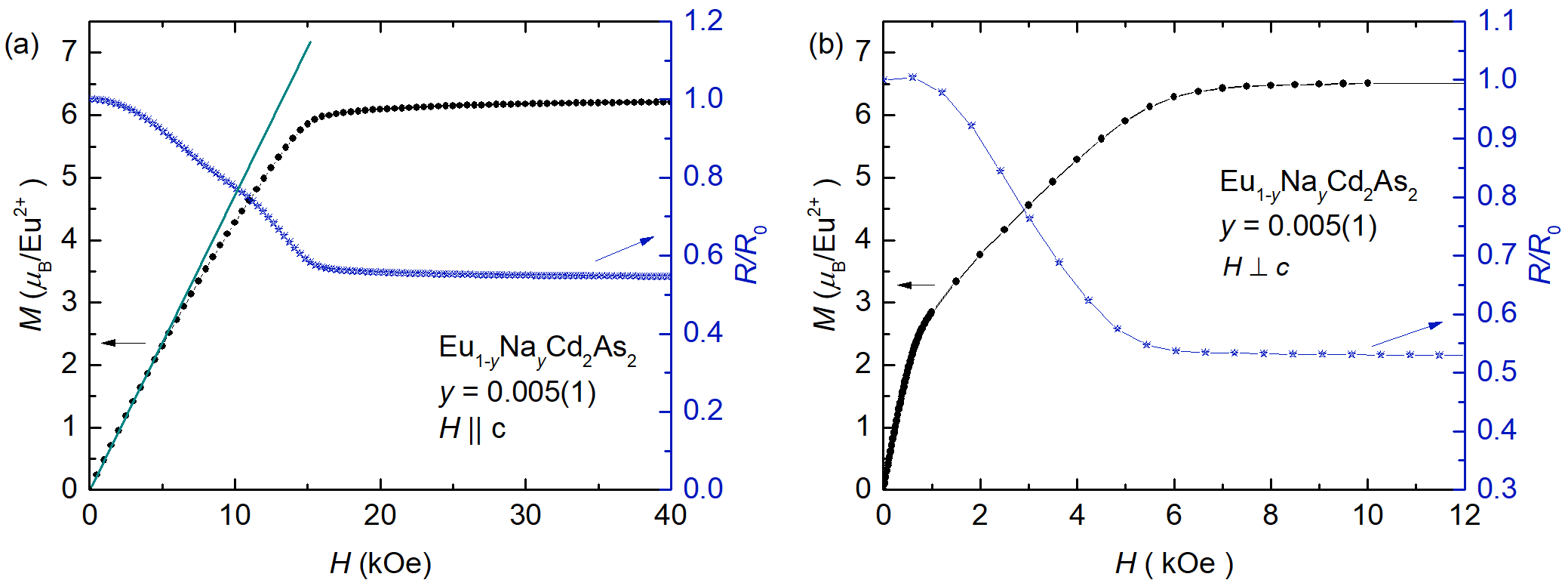}
	\caption{Magnetization as a function of field $M(H)$ measured at $T=1.8~$K and normalized resistivity as a function of field $R/R_0(H)$ measured at $T=2~$K for Na doped EuCd$_2$As$_2$ with $y=0.005(1)$. (a) Applied field $H\parallel c$. The line is drawn to show the deviation from linear behavior, corresponding to the feature in $R/R_0(H)$. (b) $H \perp c$}
	\label{Na10}
\end{figure*}

\begin{figure*}
	\includegraphics[scale=1]{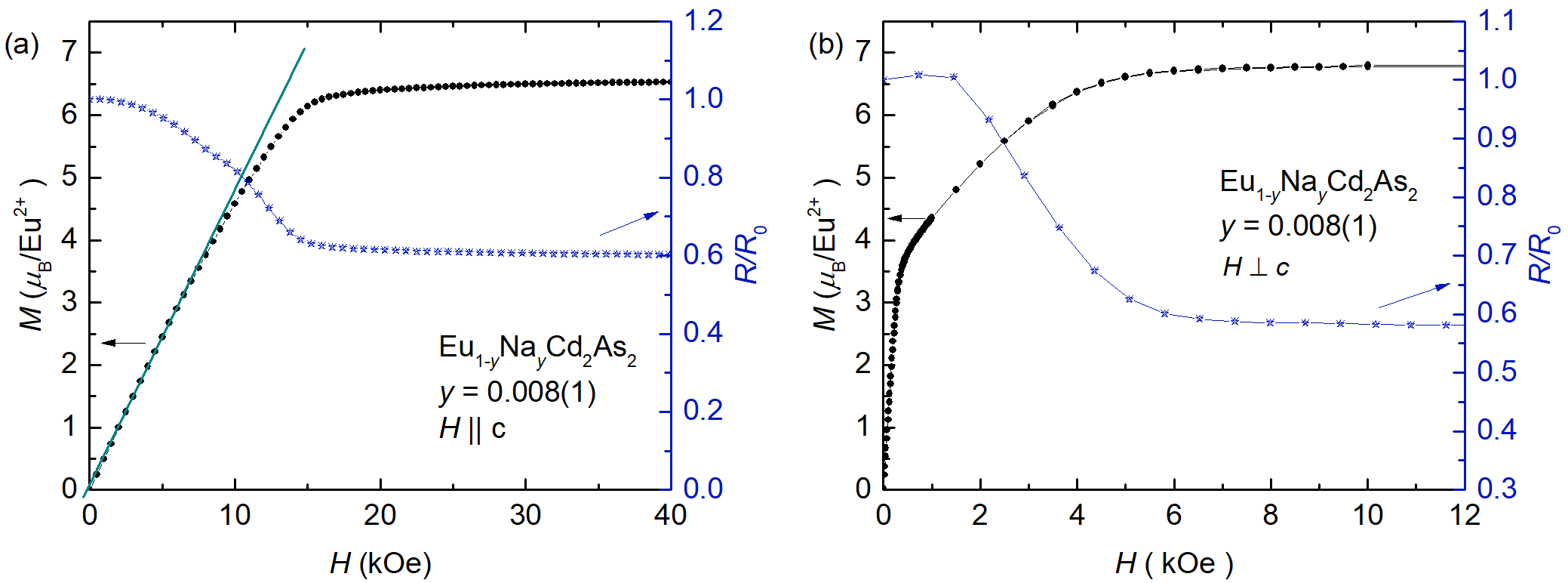}
	\caption{Magnetization as a function of field $M(H)$ measured at $T=1.8~$K and normalized resistivity as a function of field $R/R_0(H)$ measured at $T=2~$K for Ag doped EuCd$_2$As$_2$ with $y=0.008(1)$. (a) Applied field $H\parallel c$. The line is drawn to show the deviation from linear behavior, corresponding to the feature in $R/R_0(H)$. (b) $H \perp c$}
	\label{Na20}
\end{figure*}

A detailed comparison of the anisotropic $M(H)$ and $R/R_0(H)$ data, measured at $1.8~$K for magnetization and $2~$K for resistance, are shown for each sample in Figs. \ref{Ag5} - \ref{Na20}. It can be easily seen that both $M(H)$ and $R/R_0(H)$ data saturate at similar fields, showing the alignment of Eu moments with the applied filed. Additionally, one can see the feature in $R/R_0(H)$ in the intermediate fields corresponding well with the change in slope of $M(H)$. For $H \perp c$ direction, $M(H)$ shows a steep increase at low fields, followed by a slower increase at intermediate fields, and finally saturates around $5-8~$kOe. Corresponding feature in the $R/R_0(H)$ data is subtle and it shows gradual change at low - intermediate fields and eventually saturates around the same field as $M(H)$ data. On the other hand, for $H \parallel c$ direction of measurement, $R/R_0(H)$ shows a clearer/stronger feature at the intermediate fields, corresponding to the less prominent change of slope in $M(H)$ data. This is clearer in the $M(H)$ and $R/R_0(H)$ of the Na-substituted EuCd$_2$As$_2$ samples as shown in Figs. \ref{Na10} - \ref{Na20}.

\begin{figure*}
	\includegraphics[width=\linewidth]{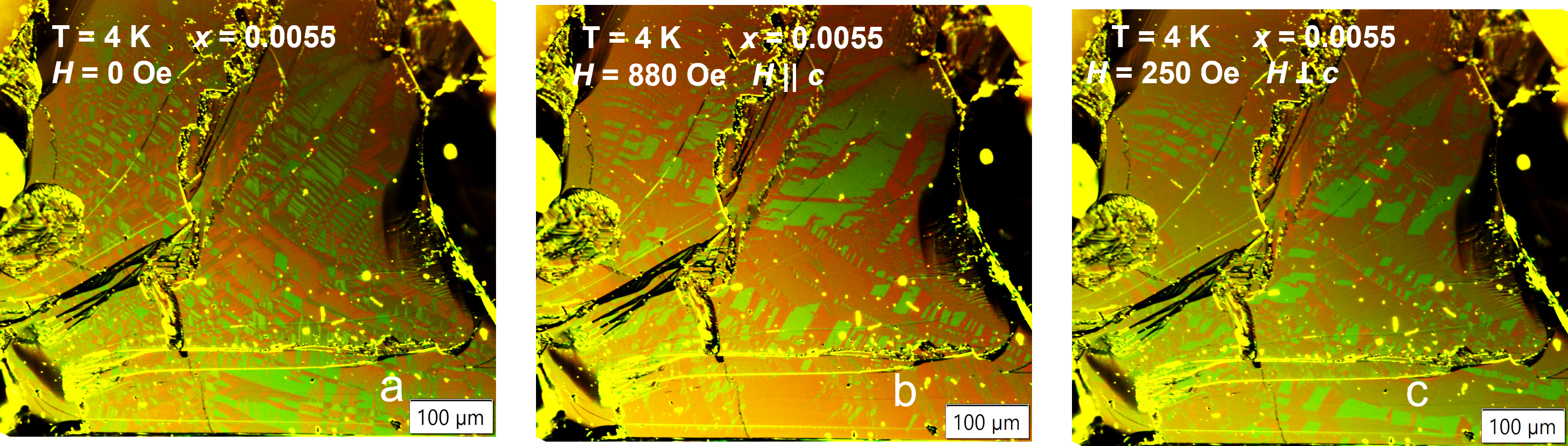}
	\caption{Magneto optical images of EuCd$_2$As$_2$, Ag substituted sample $x = 0.0055$ at $T=4~$K (a) with no applied magnetic field (b) with an applied magnetic field of $H=880~$Oe out of plane ($H\parallel c$). (c) with an applied field of $H=250~$Oe in the $ab-$plane. All three show formation of domains, which gets modified with applied filed.}
	\label{mo_sm}
\end{figure*}

Magneto optical images of $x=0.0055(3)$ Ag-substituted EuCd$_2$As$_2$ with different applied magnetic fields are shown in Fig. \ref{mo_sm}. Figure \ref{mo_sm} shows images taken at $T=4~$K with (a) zero applied field, (b) $H=880~$Oe along $H \parallel c$, and (c) $H=250~$Oe in the $ab-$plane. All three show formation of domains, which gets modified with applied filed, indicating that the domains are magnetic in nature. This also shows that the domains change with applied fields both along $c-$axis and in-plane.  

\begin{figure*}
	\includegraphics[width=\linewidth]{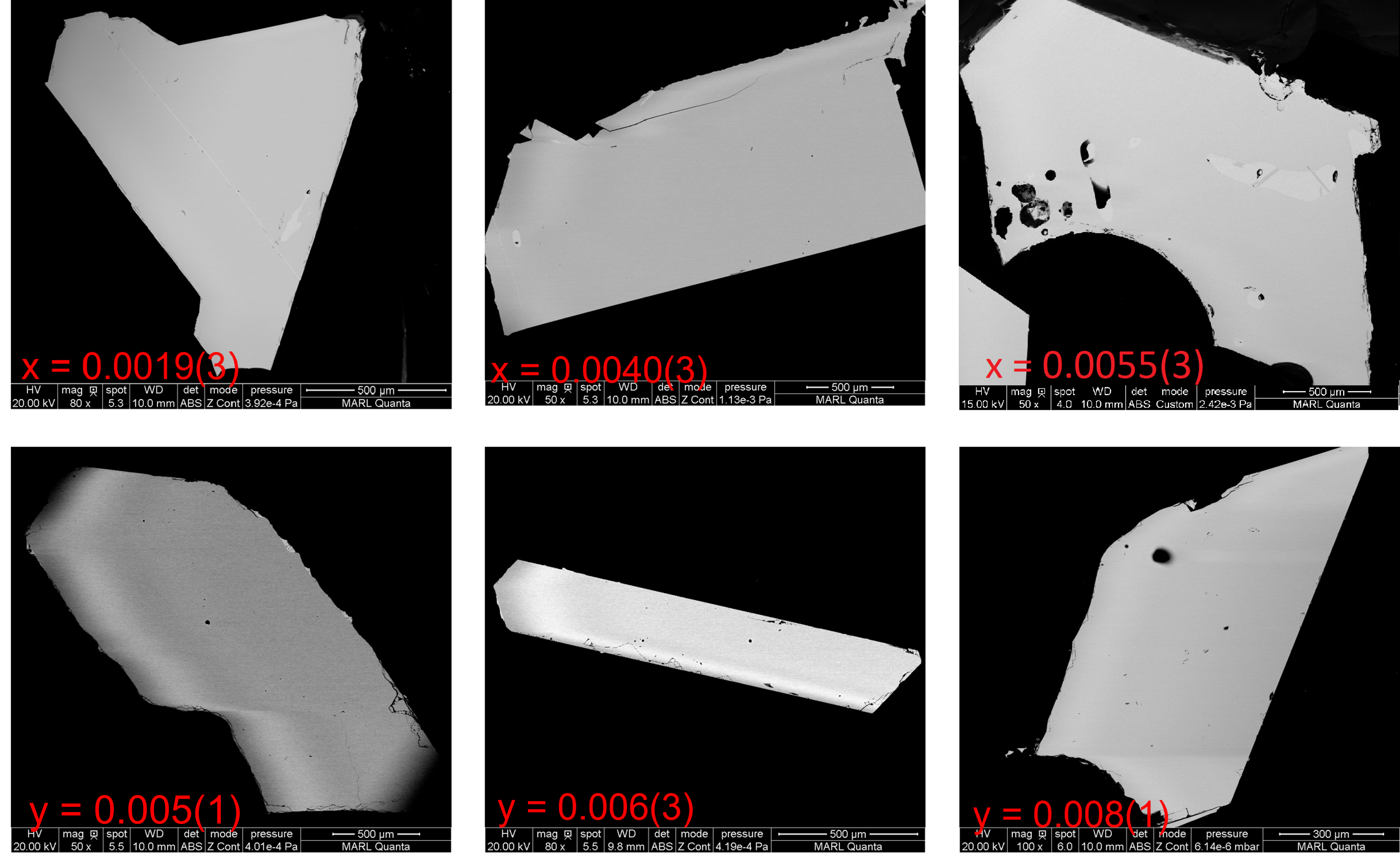}
	\caption{Upper panel shows SEM back scattering images of three Ag doped EuCd$_2$As$_2$ crystals. And lower panel shows three Na doped samples.}
	\label{sem}
\end{figure*}

Figure \ref{sem} shows the back scattering images of three Ag-substituted and three Na-substituted EuCd$_2$As$_2$ samples. There are no clear signatures of contrast indicating secondary phases, or inclusions, except for the small inclusion seen in highest doped Ag sample. 

\clearpage

\section*{References}

\bibliographystyle{apsrev}

\bibliography{EuCd2As2}

\end{document}